\documentclass[prd,preprint]{revtex4}
\usepackage[dvips]{graphicx}
\usepackage{url}

\begin{document}

\title{Classical QGP : III. The free energy}

\author{Sungtae Cho and Ismail Zahed}
\email{scho@grad.physics.sunysb.edu,
zahed@zahed.physics.sunysb.edu}
\address{Department of Physics and Astronomy \\
State University of New York, Stony Brook, NY, 11794}

\begin{abstract}
We explore further the classical QGP using methods from classical
liquids. The partition function of an ensemble of $SU(N_c)$
colored charge spheres is constructed. We analyze it using a
cumulant expansion (low density) and a loop expansion (high
temperature) after resumming the Debye screening effects. The
pertinent free energies are derived in both limits and compared to
recent molecular dynamics results.
\end{abstract}

\maketitle

\newpage

\section{Introduction}

Plasmas are statistical classical/quantum systems involving
charge constituents. Notable are electromagnetic plasmas
whereby the underlying constituents interact through long
range Coulomb fields. The simplest theory of electromagnetic
plasmas is the One Component Plasma (OCP) whereby the constituents
are like (negative) charges embedded in a uniform and neutralizing
unlike (positive) background. A number of analytical approaches
to the OCP exist in the form of diagrammatic or field theoretical
methods~\cite{hansen&mcdonald, nets&orland}. These formal approaches
form a useful theoretical corpus for understanding ionic liquids.

Ionic liquids are characterized by a short range repulsive core in
addition to the long range Coulomb potential. The core is a scalar
with a range of the order of the interparticle distance, while the
Coulomb attraction is vectorial with infinite range. Reference
liquids are described by a repulsive pair potential $w(x)=\infty$
for $x<\sigma$ and $w(x)=0$ for $x>\sigma$ where $\sigma$ is the
diameter of a hard sphere. The Coulomb potential is generally
viewed as a {\it perturbation} added to the repulsive core. Many
models of ionic liquids have been developed to accommodate
arbitrary charges and cores such as the Primitive Model (PM).
Spinoffs are the Restricted Primitive Model (RPM) with opposite
charge pairs and the same core, and the Special Primitive Model
(SPM) with arbitrary charges and the same core.

The Classical Quark Gluon Plasma (cQGP) as developed by Gelman,
Shuryak and Zahed can be regarded as an example of an SPM model in
ionic liquids, albeith with {\it non-Abelian} colored
charges~\cite{borisetal}. The nature of the core in the CQGP is
quantum mechanical and thus assumed. Detailed molecular dynamics
simulations of the cQGP~\cite{borisetal} have shown a strongly
coupled plasma for $\Gamma=V/K\approx 1$, i.e. whenever the
potential energy is of the order of the kinetic energy. The
classical and colored cQGP maybe in a liquid state at moderate
values of $\Gamma$ prompting us to use methods for classical
liquids to analyze it.

This paper is the first of a series of sequels to~\cite{borisetal}
to develop analytical methods to address the many-body dynamics of
the cQGP. In section 2, we define the partition function of the
cQGP as a classical but colored liquid. In section 3 and 4, we
work out the partition function in the low density limit through a
cumulant expansion after resumming the Debye screening effects. In
section 5, we discuss a high temperature expansion. In section 6,
we detail a diagrammatic or loop expansion for the cQGP that is
justified at high temperatures. We carry the expansion to three
loops. In section 7, we unwind the free energies both in the loop
and density expansions. In section 8, we compare the excess free
energies to recent molecular dynamics simulations. Our conclusions
are in section 9. Useful definitions and color integrations are
given in the Appendices.

\section{The grand canonical partition function for cQGP}

\renewcommand{\theequation}{II.\arabic{equation}}
\setcounter{equation}{0}

The cQGP has been defined in~\cite{borisetal}. It consists of
classical particles of like-mass $m$ with 3-position $\vec{x}_i(t)$
and 3-momentum  $\vec{p}_i(t)$ with $N_c^2-1$ adjoint colored
charges $Q^\alpha_i(t)$ and fixed $N_c-1$ Casimirs. Because of the
constraints, the color variables are Darboux's type as summarized
in the Appendix. (see also~\cite{johnson}). Particle motion in
the cQGP is treated classically through

\begin{eqnarray}
& & m\frac{dx_i^{\mu}(t)}{dt}=p_i^{\mu}(t) \nonumber \\
& & m\frac{dp_i^{\mu}(t)}{dt}=gQ^{\alpha}F_{a}^{\mu\nu}(x_i)p_{i\nu}(t) \nonumber \\
& & m\frac{dQ_i^{\alpha}(t)}{dt}=-gf^{\alpha\beta\gamma} p_i^{\mu}(t)
A^{\beta}_{\mu}(x_i)Q_i^{\gamma}(t) \label{eq2-01}
\end{eqnarray}
where we have trivially covariantized the notations. The repulsive
effects of the core is subsumed~\cite{borisetal}. $A^\mu$ is the
gauge field on the $i$th particle due to all other particles,
$F_{\alpha}^{\mu\nu}$ is its field strength, and
$f^{\alpha\beta\gamma}$ the pertinent structure constant for
$SU(N_c)$. The last relation is Wong's equation~\cite{wong}. The
magnetic contribution to the Lorentz force is suppressed by $v/c$
and dropped in the electric cQGP. cQGP simulations with both
electric and magnetic charges can be found in~\cite{liao&shuryak}.

The Hamiltonian for the cQGP reads

\begin{equation}
H=\sum_{a,i}\frac{p^2_{a,i}}{2m_a}+\frac{1}{2}\frac{g^2}{4\pi}\sum_{a,i\neq
a',j}\frac{Q^{\alpha}_{a,i}Q^{\alpha}_{a',j}}{\mid \vec
x_{a,i}-\vec x'_{a',j}\mid}+ V_{core} \label{eq2-02}
\end{equation}
where the core potential is now explicit. The double sum is over species $a$
(to be set to 3 below) and particle index $i=1,..,N$. In terms of the
color charge density

\begin{eqnarray}
\rho^{\alpha}(\vec r)=\sum_{a,i}
Q^{\alpha}_{a,i}\delta(\vec r -\vec r_{a,i})
\end{eqnarray}
the grand partition function for the cQGP is

\begin{equation}
Z=\sum_{{N_a}}\int\prod_a\frac{1}{{N_a}!}\prod_{a,i}^{N_a}(dQ_{a,i}d^{3}r_{a,i}
n_{a})\prod_{\alpha}^{N_c^2-1}\exp{\bigg(-\frac{1}{2}\beta\int
d^{3}r d^{3}r' \rho^{\alpha}(\vec r)v(\vec r-\vec
r')\rho^{\alpha}(\vec r')\bigg)} \label{eq2-03}
\end{equation}
with

\begin{equation}
v(\vec r-\vec r')=\frac{g^2}{4\pi}\frac{1}{\mid \vec r-\vec r'
\mid} \label{eq2-04}
\end{equation}
and the free particle density

\begin{equation}
n_{a}=g_a \frac{1}{{\Lambda_a}^3}e^{\beta \mu_{a}} \label{eq2-05}\,\,.
\end{equation}

\noindent $g_a$ is the degeneracy factor and $\mu_{a}^{\alpha}$
the chemical potential of species $a$. The factor of
${\Lambda}_{a}^{-3}$ is the thermal wavelength resulting from the
momentum integral over phase space. The repulsive core potential
is set to be

\begin{equation}
w(\vec r-\vec r') = \left\{ \begin{array}{ll} \infty
 & (\mid \vec r-\vec r' \mid < \sigma) \nonumber \\
0  & (\mid \vec r-\vec r' \mid > \sigma) \end{array} \right.
\label{eq2-06}
\end{equation}

with a core size $\sigma$ to prevent classical collapse. In
reality, the core emerges from Coulomb repulsion between
like-particles or the quantum uncertainty repulsion between
unlike/like particles. With this in mind, we can rewrite
(\ref{eq2-03}) in the form

\begin{eqnarray}
\lefteqn{ Z=\sum_{{N_a}}\int\prod_a\frac{1}{{N_a}!}
\prod_{a,i}^{N_a}(dQ_{a,i}d^{3}r_{a,i} n_{a})
\prod_{\alpha}^{N_c^2-1} } \nonumber \\
& \times \exp{\bigg(-\frac{1}{2}\beta\int d^{3}r d^{3}r'
\rho^{\alpha}(\vec r)v(\vec r-\vec r') \rho^{\alpha}(\vec
r')\bigg)} e^{\frac{1}{2}w_0} \exp{\bigg(-\frac{1}{2}\int d^{3}r
d^{3}r' \rho(\vec r)w(\vec r-\vec r')\rho(\vec r')\bigg)}
\nonumber \\
\label{eq2-07}
\end{eqnarray}
with the number density $\rho(\vec r)=\sum_{a,i}\delta(\vec r -\vec r_{a,i})$.
$w_0=w(0)$ is identified with the self-energy.

The repulsive core (\ref{eq2-06}) acts like the potential between
two hard spheres with diameter $R=\sigma$. Thus, we may treat each
particle classically as a rigid sphere of diameter $R$ with uniform
color charge $Q^\alpha$ on the surface,

\begin{eqnarray}
Q\,q(\vec{x}) =\frac{Q}{\pi R^2}\delta (|\vec{x}|-R/2)
\end{eqnarray}
The repulsive core (\ref{eq2-06}) is now generated classically as
the Coulomb repulsion between the rigid
spheres~\cite{caillol&raimbault}. For two spheres located at
$\vec{r}$ and $\vec{r}'$ this is

\begin{equation}
W(\vec r-\vec r')=\int d\vec x d\vec y q(\mid\vec r-\vec
x\mid)\frac{1}{\mid \vec x-\vec y \mid}q(\mid\vec r'-\vec y\mid)\,\,.
\label{eq2-08}
\end{equation}
The grand canonical partition function follows in the form

\begin{eqnarray}
\lefteqn{Z=\sum_{{N_a}}\int\prod_a\frac{1}{{N_a}!}\prod_{a,i}^{N_a}
(dQ_{a,i}d^{3}r_{a,i} n_{a}) e^{\gamma(N_c^2-1)W_0} e^{-v_{HS}} }
\nonumber \\
& \times \exp\Big(-\frac{\beta}{2}\frac{g^2}{4\pi}
\sum_{\alpha}^{N_c^2-1} \int d^3r d^3r'\rho^{\alpha}(\vec r)
W(\vec r-\vec r')\rho^{\alpha}(\vec r')\Big) \label{eq2-09}
\end{eqnarray}

where the Coulomb self energy is now $W_0=W(\vec 0)/2$ and $\gamma_a$

\begin{equation}
\gamma_a=\beta\frac{g^2}{4\pi}\frac{1}{N_c^2-1}\sum_{\alpha}{Q_a^{\alpha}}^2
\label{eq2-10} \end{equation}

\section{Liquid of Colored Hard Spheres}

\renewcommand{\theequation}{III.\arabic{equation}}
\setcounter{equation}{0}

To analyze (\ref{eq2-09}) we first consider a dilute cQGP
with $N_c$ colors. Performing $(N_c^2-1)$ times the Sine-Gordon
transform on (\ref{eq2-09}) yields

\begin{equation}
\exp\Big(-\frac{\beta}{2}\frac{g^2}{4\pi}\sum_{\alpha}^{N_c^2-1}\int
d\vec r d\vec r' \rho^{\alpha}(\vec r)W(\vec r-\vec
r')\rho^{\alpha}(\vec r')\Big) =\langle\exp\Big(
\sum_{\alpha}^{N_c^2-1} i(\beta\frac{g^2}{4\pi})^{1/2}\int d\vec r
\rho^{\alpha}(r)\phi^{\alpha}(r)\Big)\rangle_{W}\label{eq3-01}
\end{equation}
where the averaging on the RHS (right hand side) is carried using the measure

\begin{equation}
\langle\cdots\rangle_{W}=\frac{\int[\prod_{\alpha}^{N_c^2-1}d\phi^{\alpha}](\cdots)
\exp\Big(-\frac{1}{2}\sum_{\alpha}^{N_c^2-1}\int d\vec r d\vec r'
\phi^{\alpha}(\vec r) W^{-1}(\vec r-\vec r')\phi^{\alpha}(\vec
r')\Big)} {\int[\prod_{\alpha}^{N_c^2-1}d\phi^{\alpha}]
\exp\Big(-\frac{1}{2} \sum_{\alpha}^{N_c^2-1} \int d\vec r d\vec
r' \phi^{\alpha}(\vec r) W^{-1}(\vec r-\vec r')\phi^{\alpha}(\vec
r')\Big)} \label{eq3-02}
\end{equation}
To proceed further, we define

\begin{equation}
\tilde{\nu}_{a}=\beta\tilde{\mu}_{a}
=\beta{\mu}_{a}+(N_c^2-1)\gamma W_0 \qquad \tilde{n}_{a}=n_a
e^{(N_c^2-1)\gamma W_0} e^{i(\beta\frac{g^2}{4\pi})^{\frac{1}{2}}
\sum_{\alpha}^{N_c^2-1} \sum_{a,i} Q_{a,i}^{\alpha}
\phi^{\alpha}(\vec r_{a,i})} \label{eq3-03}
\end{equation}
as the renormalized chemical potential and species densities
respectively. In terms of which the grand canonical partition
function is now

\begin{equation}
Z=\sum_{{N_a}}\int\prod_a\frac{1}{{N_a}!}\prod_{a,i}^{N_a}(dQ_{a,i}d^{3}r_{a,i}
\langle {\tilde{n}}_{a}\rangle_{W})e^{-v_{HS}}\equiv\langle Z_{HS}
\rangle_{W} \label{eq3-04}
\end{equation}
$e^{-v_{HS}}$ is short for the Gaussian measure in (\ref{eq3-02}).
(\ref{eq3-02}) can be Taylor expanded around the mean density $n_a$
(see below). The result is

\begin{eqnarray}
\ln\bigg(\frac{Z}{Z_{HS}}\bigg) & = & \sum_{n=1}\frac{1}{n!}
\sum_{a_1,a_2,\cdots, a_n} \int
(d^{3}r_{a_1}dQ_{a_1})\cdots(d^{3}r_{a_n}dQ_{a_n})\nonumber \\
& \times & \frac{\delta^n \ln Z}{\delta\tilde {n}_{a_1}(\vec
r_{a_1})\cdots\delta\tilde{n}_{a_n}(\vec r_{a_n})}
\bigg\vert_{\tilde{n}_{a_i}(\vec r_{a_i})=n_{a}}
\prod_{i=1}^{n}(\tilde{n}_{a_i}(\vec r_{a_i})-n_a) \label{eq3-05}
\end{eqnarray}
If we were to assume that the species chemical potentials are
all the same $\mu_{a_1}=\cdots=\mu_{a_n}=\mu_{a}$, then in
the free particle case

\begin{equation}
\mu_{0}=\mu_a+\frac{1}{\beta}\ln(3), \qquad n_0=3n_a=3\lambda
\label{eq3-06}
\end{equation}
We have set the particle species to 3 to account classically for
quarks, anti-quarks and gluons all of equal constituent thermal
mass $m$ for simplicity. In terms of (\ref{eq3-06}) the integrand
in (\ref{eq3-03}) can be identified with the classical correlation
function $h_0^{(n)}$ of a liquid of hard
spheres~\cite{caillol&raimbault}

\begin{equation}
{n_a}^n \frac{\delta^n \ln Z}{\delta\tilde {n}_{a_1}(\vec
r_{a_1})\cdots\delta\tilde{n}_{a_n}(\vec r_{a_n})}
\bigg\vert_{\tilde{n}_{a_i}(\vec
r_{a_i})=n_{a}}=\frac{\rho_{0}'^{n}}{3^n}h_{0}^{(n)}(\vec
r_{a_1},\cdots \vec r_{a_n})=\rho_{0}^{n}h_{0}^{(n)}(\vec
r_{a_1},\cdots \vec r_{a_n}) \label{eq3-07}
\end{equation}
with the number density

\begin{equation}
\rho_{0}'=\frac{1}{V}\frac{\partial \ln Z_{HS} }{\partial \nu_{0}}
\label{eq3-08}
\end{equation}
Thus, the partition function for a colored liquid of hard spheres now read

\begin{equation}
\ln\bigg(\frac{Z}{Z_{HS}}\bigg)=\langle \exp(-U[\phi^{\alpha}])
\rangle_W=\mathcal{N}_W^{-1}\int[\prod_{\alpha}d\phi^{\alpha}]
\exp(-S[\phi^{\alpha}]) \label{eq3-09}
\end{equation}
with

\begin{eqnarray}
\lefteqn{\qquad \mathcal{N}_W=\int[\prod_{\alpha}d\phi^{\alpha}]
\exp\Big(-\frac{1}{2} \sum_{\alpha}^{N_c^2-1} \int d\vec r d\vec
r' \phi^{\alpha}(\vec r) W^{-1}(\vec r-\vec
r')\phi^{\alpha}(\vec r')\Big) } \nonumber \\
& & S[\phi^{\alpha}]=\frac{1}{2} \sum_{\alpha}^{N_c^2-1} \int
d\vec r d\vec r' \phi^{\alpha}(\vec r) W^{-1}(\vec r-\vec
r')\phi^{\alpha}(\vec r')
+ \sum_{n=1}U_n[\phi^{\alpha}] \nonumber \\
& & U_n[\phi^{\alpha}]=-\frac{\rho_{0}^n}{n!}\sum_{a_1,a_2,\cdots,
a_n}\int (d^{3}r_{a_1}dQ_{a_1})\cdots(d^{3}r_{a_n}dQ_{a_n})
h_{0}^{(n)}\prod_{i=1}^{n}(\frac{\tilde{n}_{a_i}(\vec
r_{a_i})}{n_a}-1) \label{eq3-10}
\end{eqnarray}

\section{Cumulant Expansion}

\renewcommand{\theequation}{IV.\arabic{equation}}
\setcounter{equation}{0}

To perform a cumulant expansion at low density, we add and
subtract the Hartree part in $U[\phi]$ through

\begin{equation}
U[\phi^{\alpha}]=(U[\phi^{\alpha}]-U_{0}[\phi^{\alpha}])
+U_{0}[\phi^{\alpha}] \label{eq3-11}
\end{equation}
with

\begin{equation}
U_{0}[\phi^{\alpha}]=\frac{1}{2}\rho_{0}\sum_{a}\gamma_a\sum_{\alpha}
\int d\vec r {\phi^{\alpha}(\vec r)}^2\,\,, \label{eq3-12}
\end{equation}
yields the identity

\begin{equation}
\langle \exp(-U[\phi^{\alpha}])
\rangle_W=\frac{\mathcal{N}_X}{\mathcal{N}_W}\langle
\exp(-U[\phi^{\alpha}]) \rangle_X \label{eq3-13}
\end{equation}
with X$^{-1}=$W$^{-1}+\rho_{0}\sum_a\gamma_a${\bf 1}.
The normalizations in (\ref{eq3-12}) is explicitly

\begin{equation}
\frac{\mathcal{N}_X}{\mathcal{N}_W}=\exp{\Big(-\frac{V}{2}(N_c^2-1)\int
\frac{d\vec q}{(2\pi)^3}\ln( 1+\rho_{0}\sum_{a}\gamma_a
\tilde{W}(\vec q)) \Big)} \label{eq3-14}
\end{equation}
The Fourier transform of the core $\tilde{W}(q)$ is related
to the Fourier transform $\tilde{X}(q)$ as

\begin{equation}
\tilde{X}(q)=\frac{\tilde{W}(q)}{1+\rho_{0}\gamma\tilde{W}(\vec
q)}=\frac{\sin^2(qR/2)}{(qR/2)^2}\frac{4\pi}{q^2+\kappa_{0}^2
\frac{\sin^2(qR/2)}{(qR/2)^2}} \label{eq3-15}
\end{equation}
with

\begin{equation}
\kappa_{0}^2 =4\pi \sum_{a}\gamma_a\rho_{0}=\beta g^2\rho_{0}
\frac{1}{N_c^2-1}\sum_a\sum_{\alpha}{Q_a^{\alpha}}^2
\label{eq3-16}
\end{equation}

All in all the partition function for a colored liquid of hard
spheres with their color charges smeared on the surface is

\begin{equation}
-\frac{\ln{Z}}{V}=-\frac{\ln{Z_{HS}}}{V}+\frac{1}{2}(N_c^2-1) \int
\frac{d^3\vec q}{(2\pi)^3}\ln( 1+\rho_{0}\sum_a\gamma_a
\tilde{W}(\vec q))-\frac{1}{V}\sum_{n=1}\frac{(-1)^n}{n!}
\langle\mathcal{H}^n[\phi^{\alpha}]\rangle_{X,c} \label{eq3-17}
\end{equation}
with $\mathcal{H}[\phi^{\alpha}]=U[\phi^{\alpha}]
-U_0[\phi^{\alpha}]$ and $\langle\cdots\rangle_{X,c}$ denotes a
cumulant average. At low density, the grand partition function
is dominated by the first cumulants

\begin{equation}
-\frac{\ln{Z}}{V}=-\frac{\ln{Z_{HS}}}{V}+w_1+w_2+\mathcal{O}(\rho_0^3)
\label{eq3-18} \end{equation}
with

\begin{eqnarray}
\lefteqn{\quad w_1= \frac{1}{2}(N_c^2-1) \int \frac{d\vec
q}{(2\pi)^3}\ln( 1+\rho_{0}\sum_a\gamma_a \tilde{W}(\vec q)) }\nonumber \\
& & w_2 = \frac{1}{V}\langle\mathcal{H}[\phi^{\alpha}]\rangle_{X}-
\frac{\langle\mathcal{H}^2[\phi^{\alpha}]\rangle_{X}-
\langle\mathcal{H}[\phi^{\alpha}]\rangle_{X}^2}{2V} \nonumber \\
& & \mathcal{H}=U_1[\phi^{\alpha}]+U_2[\phi^{\alpha}]-
U_0[\phi^{\alpha}]\label{eq3-19}
\end{eqnarray}

The leading or zeroth order cumulant is known from hard sphere
liquid analysis~\cite{hansen&mcdonald}

\begin{equation}
-\frac{\ln{Z_{HS}}}{V}=-\rho_{0}'-\frac{2\pi}{3}\rho_{0}'^2\sigma^3-
\frac{5}{18}\pi^2\rho_{0}'^3\sigma^6+\mathcal{O}(\rho_{0}'^4)
\label{eq3-20}
\end{equation}
with $\rho'_0=3\rho_0$. The first cumulant reads

\begin{eqnarray}
\lefteqn{ w_1-(N_c^2-1) \rho_{0} \sum_a\gamma_a W(0)}  \nonumber \\
& &=-(N_c^2-1)\frac{2}{3}\sqrt{\pi}{\rho_{0}}^{3/2}
(\sum_a\gamma_a)^{3/2}+ (N_c^2-1)\frac{7}{15} \pi{\rho_{0}}^2
(\sum_a\gamma_a)^2R \nonumber \\
& &-(N_c^2-1)\frac{1}{3}{\pi}^{3/2}{\rho_{0}}^{5/2}
(\sum_a\gamma_a)^{5/2}R^2 +\mathcal{O}(\rho_{0}^3)\,\,.\label{eq3-21}
\end{eqnarray}
The Gaussian averaging in the second cumulant can be carried
out. The result is

\begin{eqnarray}
\lefteqn{w_2=-(N_c^2-1)\frac{\rho_{0}}{R}\sum_a\gamma_a-\frac{1}{2}
(N_c^2-1)\rho_{0}\sum_a\triangle_{a0} +\rho_{0}\bigg(3-\sum_a
e^{-\frac{1}{2}(N_c^2-1)\triangle_{a0}}\bigg) } \nonumber \\
& & -\frac{\rho_{0}^2}{4}\bigg(\sum_a\gamma_a (1-e^{-\frac{1}{2}
(N_c^2-1)\triangle_{a0}})\bigg)^2 \int d\vec r X^2(r)
-\frac{1}{2}\rho_{0}^2 \Big(3-\sum_a e^{-\frac{1}{2}(N_c^2-1)
\triangle_{a0}}\Big)^2 \tilde{h}_0(0)\nonumber \\
& & -\frac{1}{4}\rho_{0}^2(N_c^2-1) \Big(\sum_a\gamma_a
e^{-\frac{1}{2}(N_c^2-1)\triangle_{a0}}\Big)^2 \int d\vec r
X^2(r)h_0^{(2)}(r) \nonumber \\
& & -\frac{1}{2}\rho_{0}^2\sum_{a,a'}\bigg(e^{-\frac{1}{2}
(N_c^2-1) \triangle_{a0}}e^{-\frac{1}{2}(N_c^2-1)\triangle_{a'0}}
\int d\vec r dQ_a dQ_{a'}\nonumber \\
& & \times \Big(e^{-\beta \frac{g^2}{4\pi}X(r)
\sum_{\alpha}Q_a^{\alpha} Q_{a'}^{\alpha}}-1-\frac{1}{2}(N_c^2-1)
\gamma_a\gamma_{a'}X^2(r)\Big)(h_0^{(2)}(r)+1) \bigg)
\label{eq3-22}
\end{eqnarray}
with $\triangle_{a0}=\gamma_a X(0)-2\gamma_a/R$ and $\tilde{h}_{0}(0)$
the 3D Fourier transform of $h_{0}^{(2)}(r)$,

\begin{equation}
\triangle_{a0}=\gamma_a X(0)-2\frac{\gamma_a}{R}=\gamma_a\bigg(
-\kappa_{0}+\frac{7}{15} \kappa_{0}^2 R -\frac{5}{24} \kappa_{0}^3
R^2+\mathcal{O}(\kappa_{0}^4) \bigg) \label{eq3-23}
\end{equation}
and

\begin{eqnarray}
\lefteqn{\quad \gamma_a
X(r)=2\frac{\gamma_a}{R}-\frac{\gamma_a}{R^2}r-\gamma_a \kappa_{0}
+\mathcal{O}(\kappa_{0}^2) \qquad (r<\sigma) } \nonumber \\
& & \gamma_a X(r)=\gamma_a \frac{\sin^2(\kappa_{0}
R/2)}{(\kappa_{0} R/2)^2} \frac{1}{r}e^{(-\kappa_{0}
r)}+\mathcal{O}(\kappa_{0}^2) (r>\sigma) \label{eq3-24}
\end{eqnarray}

Collecting all the results up to the second cumulant in powers of the
original activity $n_a (=\lambda)$ we have for $N_c=2$

\begin{equation}
-\frac{\sigma^3\ln{Z}}{V}=-3\tilde{\lambda}-2\sqrt{\pi}
\tilde{\lambda}^{3/2}(\sum_a\epsilon_a)^{3/2}-\tilde{\lambda}^2
\Big(\frac{9}{2}\pi \sum_a\epsilon_a \sum_a\epsilon_a^2+{B_2}'
\Big) + \mathcal{O}(\tilde{\lambda}^{5/2}) \label{eq3-25}
\end{equation}
up to order $\lambda^{5/2}$, with

\begin{equation}
{B_2}'=\frac{\pi}{24\epsilon}\sum_{a,a'}\bigg(54\epsilon^4
\int_{0}^{3\epsilon} \frac{\sinh t }{t}dt +
e^{3\epsilon}(-2-2\epsilon-3\epsilon^2-9\epsilon^3) +
e^{-3\epsilon}(2-2\epsilon+3\epsilon^2-9\epsilon^3) \bigg)
\label{eq3-26}
\end{equation}

\noindent where $\epsilon_a=\frac{\gamma_a}{\sigma}$,
$\epsilon=\sqrt{\epsilon_a \epsilon_{a'}}$ and
$\tilde{\lambda}=\sigma^3\lambda$. We have set $h_0^{(2)}(r)=-1$
for $r<\sigma$ and $h_0^{(2)}(r)=0$ otherwise, and used

\begin{equation}
\rho_0=\frac{\rho_0'}{3}=\frac{n_0}{3} \frac{\partial}{\partial
n_0}\Big(\frac{\ln{Z_{HS}}} {V}\Big)=\frac{n_0}{3}-\frac{1}{3}
\frac{4\pi}{3} \sigma^3n_0^2+\mathcal{O}(n_0^3)=\lambda
-3\frac{4\pi}{3}\sigma^3 \lambda^2+\mathcal{O}(\lambda^3)
\label{eq3-27}
\end{equation}

The correction of order $\lambda^{5/2}$ is straightforward
but tedious. For any $N_c$ we obtain

\begin{eqnarray}
& & -(N_c^2-1)^3\frac{\pi^{\frac{3}{2}}}{4}
(\sum_a\gamma_a)^{\frac{1}{2}} (\sum_a\gamma_a^2)^2
-(N_c^2-1)^3\frac{\pi^{\frac{3}{2}}}{6} \sum_a\gamma_a^3
(\sum_a\gamma_a)^{\frac{3}{2}}  \nonumber \\
& & +(N_c^2-1)(\sum_{a}\gamma_{a})^2
(\sum_{a}\gamma_a)^{\frac{1}{2}}2\pi^{\frac{3}{2}} \sigma^2
-\frac{1}{2} (N_c^2-1)^2 (\sum_{a}\gamma_{a}^2) (\sum_{a}\gamma_a
)^{\frac{3}{2}}2\pi^{\frac{3}{2}} \sigma \nonumber \\
& & - \sqrt{\pi}(\sum_{a}\gamma_a)^{1/2}\sum_{a,a'}
\int_{\sigma}^{\infty} d^3\vec r \bigg( \frac{r}{\gamma_{aa'}
(N_c^2-1)\frac{1}{r}} \sinh(\gamma_{aa'}(N_c^2-1)\frac{1}{r})
-r\cosh(\gamma_{aa'}(N_c^2-1)\frac{1}{r}) \nonumber \\
& & + \frac{(\gamma_{a}+\gamma_{a'0})} {2\gamma_{aa'}\frac{1}{r}}
\sinh(\gamma_{aa'}(N_c^2-1)\frac{1}{r})
+\frac{1}{2}(N_c^2-1)(\gamma_a+\gamma_{a'})
+\gamma_a\gamma_{a'}(N_c^2-1)\frac{1}{r} \nonumber \\
& & -\frac{1}{4}(\gamma_a+\gamma_{a'})\gamma_a
\gamma_{a'}(N_c^2-1)^2\frac{1}{r^2} \bigg) \label{eq3-28}
\end{eqnarray}
The left out integration is divergent. For $N_c=2$, the result
is logarithmically divergent

\begin{equation}
\frac{54}{5}\pi^{\frac{3}{2}}(\sum_a \gamma_a)^{\frac{1}{2}}
(\sum_a \gamma_a^2)^2\ln(\frac{\Lambda}{\sigma}) \label{eq3-29}
\end{equation}
where $\Lambda$ is an infrared cutoff in the radial direction.
This result is in total agreement with the result obtained earlier
in~\cite{borisetal2} using a method that did not account for hard
spheres and the smearing of the color charge. For arbitrary $N_c$
the order $\lambda^{5/2}$ correction is

\begin{eqnarray}
& & -(N_c^2-1)^3\frac{\pi^{\frac{3}{2}}}{4}
(\sum_a\gamma_a)^{\frac{1}{2}} (\sum_a\gamma_a^2)^2
-(N_c^2-1)^3\frac{\pi^{\frac{3}{2}}}{6} \sum_a\gamma_a^3
(\sum_a\gamma_a)^{\frac{3}{2}}  \nonumber \\
& & +(N_c^2-1)(\sum_{a}\gamma_{a})^2
(\sum_{a}\gamma_a)^{\frac{1}{2}}2\pi^{\frac{3}{2}} \sigma^2
-\frac{1}{2} (N_c^2-1)^2 (\sum_{a}\gamma_{a}^2) (\sum_{a}\gamma_a
)^{\frac{3}{2}}2\pi^{\frac{3}{2}} \sigma \nonumber \\
& & + \frac{54}{5}\pi^{\frac{3}{2}}(\sum_a \gamma_a)^{\frac{1}{2}}
(\sum_a \gamma_a^2)^2\ln(\frac{\Lambda}{\sigma}) \label{eq3-30}
\end{eqnarray}

\section{High Temperature Expansion}

\renewcommand{\theequation}{V.\arabic{equation}}
\setcounter{equation}{0}


Another useful way to analyze (\ref{eq3-09}-\ref{eq3-10}) is through
a high temperature expansion which parallels the cumulant (virial)
expansion. For that we order the last part of $U_n[\phi^{\alpha}]$ in
(\ref{eq3-10}) in powers of $\beta$ as in~\cite{raimbault&caillol}.
For that we define $\gamma'=\beta{g^2}/{4\pi}$ so that

\begin{eqnarray}
\frac{\tilde{n}_{a_i}(\vec r_{a_i})}{n_a}-1 &=&
\gamma_a(N_c^2-1)\frac{1}{R}+i{\gamma'}^{\frac{1}{2}}\sum_{\alpha}
Q_a^{\alpha}\phi^{\alpha}(\vec r) -\gamma'\frac{1}{2}
\sum_{\alpha, \alpha'}Q_a^{\alpha}Q_a^{\alpha'}\phi^{\alpha}(\vec
r)\phi^{\alpha'}(\vec r)\nonumber \\
&+& \gamma_a^2\frac{1}{2}(N_c^2-1)^2
\frac{1}{R^2}-\gamma_a\gamma'\frac{1}{2}(N_c^2-1)\frac{1}{R}
\sum_{\alpha, \alpha'}Q_a^{\alpha}Q_a^{\alpha'}\phi^{\alpha}(\vec
r)\phi^{\alpha'}(\vec r) \nonumber \\
&+& \frac{1}{3!}i^3{\gamma'}^{\frac{3}{2}}\sum_{\alpha,\alpha'
,\alpha''} Q_a^{\alpha}Q_a^{\alpha'}Q_a^{\alpha''}
\phi^{\alpha}(\vec r)\phi^{\alpha'}(\vec r)\phi^{\alpha''}(\vec r)
\nonumber \\
&+& \frac{1}{4!} {\gamma'}^2 \sum_{\alpha,\alpha',\alpha''
,\alpha'''} Q_a^{\alpha}Q_a^{\alpha'}Q_a^{\alpha''}
Q_a^{\alpha'''} \phi^{\alpha}(\vec r)\phi^{\alpha'}(\vec
r)\phi^{\alpha''}(\vec r)\phi^{\alpha'''}(\vec r)
+\mathcal{O}({\beta}^{\frac{5}{2}}) \label{eq3-31}
\end{eqnarray}
The rules for color averaging are summarized in Appendix B. For
instance, using (\ref{eqc07}) and (\ref{eqc17}) we have

\begin{equation}
\frac{1}{V}\langle U_1[\phi^{\alpha}]\rangle_X
=\frac{1}{2}\rho_{0} \sum_a\triangle_{a0}(N_c^2-1)
-\frac{1}{8}\rho_{0}\sum_a\triangle_{a0}^2(N_c^2-1)^2
+\mathcal{O}({\gamma}^{3}) \label{eq3-32}
\end{equation}
in leading order. In particular, the second cumulant at high
temperature is

\begin{eqnarray}
\lefteqn{ w_2 = - (N_c^2-1)\frac{1}{R}\sum_a\gamma_a\rho_{0}
-\frac{1}{8}(N_c^2-1)^2{\rho_{0}}^2\tilde{h}_0(0)
(\sum_a\triangle_{a0})^2 } \nonumber \\
& & -\frac{1}{8} \rho_{0} (\sum_a\triangle_{a0})^2(N_c^2-1)^2 -
\frac{1}{4}(N_c^2-1){\rho_{0}}^2 (\sum_a\gamma_a)^2\int d^3\vec
r{h}_{0}^{(2)}(r)X^2(\vec r) \nonumber \\
& & +\frac{1}{2}\frac{1}{3!}{\rho_{0}}^2\gamma'^3
\frac{q_3}{4}(\sum_a)^2 \int d^3\vec r({h}_{0}^{(2)}(r)+1)X^3(\vec
r)+ \mathcal{O}(\beta^{\frac{7}{2}})\,\,. \label{eq3-33}
\end{eqnarray}
The last term contributes only for $N_c>2$ with $q_3$ being
the cubic Casimir say for SU(3) as detailed in Appendix B.


After collecting all terms and replacing $X(r)$ with
$\frac{1}{r}e^{-\kappa_0 r}$ in the limit $R\to0$, we finally
obtain

\begin{eqnarray}
-\frac{\ln{Z}}{V} &=&-\frac{\ln{Z_{HS}}}{V} -\frac{2\sqrt{\pi}}{3}
(N_c^2-1) \rho_{0}^{3/2}(\sum_a\gamma_a)^{3/2}\nonumber \\
&-&\frac{1}{4}(N_c^2-1) \rho_{0}^2(\sum_a\gamma_a)^2\int d^3\vec
r{h}_{0}^{(2)}(r) \frac{1}{r^2}+\pi^{\frac{1}{2}}(N_c^2-1)
{\rho_{0}}^{\frac{5}{2}} (\sum_a\gamma_a)^{\frac{5}{2}} \int
d^3\vec r{h}_{0}^{(2)}(r)\frac{1}{r}\nonumber \\
&-&\frac{1}{2}\pi(N_c^2-1)^2\rho_{0}^3(\sum_a\gamma_a)^3\int
d^3\vec r{h}_{0}^{(2)}(r)-2\pi(N_c^2-1)\rho_{0}^3
(\sum_a\gamma_a)^3\int d^3\vec r{h}_{0}^{(2)}(r) \nonumber \\
&-&\frac{\pi}{2}(N_c^2-1)^2\rho_{0}^2(\sum_a\gamma_a^2)
(\sum_a\gamma_a) \nonumber \\
&+&{\rho_{0}}^{2} \frac{1}{2}\frac{1}{3!} \Big(
\frac{g^2}{4\pi}\Big)^3 (\sum_{a})^2 \beta^3\frac{q_3}{4}\int
d^3\vec r({h}_{0}^{(2)}(r)+1)
\frac{1}{r^3}e^{-3(4\pi\rho_{0}(\sum_a\gamma_a))^{\frac{1}{2}}r}
+\mathcal{O}(\beta^{\frac{7}{2}})\,\,. \label{eq3-34}
\end{eqnarray}
Below we show that this result yields a high temperature free
energy for the cQGP that is identical to the one following from
the loop expansion with an infinite core (\ref{eq2-06}) with
$h_0^{(2)}(r)=-\theta(\sigma-r)$.

\section{Loop Expansion}

\renewcommand{\theequation}{VI.\arabic{equation}}
\setcounter{equation}{0}

The loop expansion of (\ref{eq2-07}) is best captured by
reorganizing the expansion around the Debye-screened
solution. This expansion is identical with the high temperature expansion
of the hard sphere liquid in the limit of zero size spheres. The finite
size case will be derived by inspection. With this in mind,  we can perform
and perform the Hubbard-Stratonovitch transform on the
colored interaction part of (\ref{eq2-07}),

\begin{eqnarray}
\lefteqn{\exp{\bigg(-\frac{1}{2}\beta\int d\vec r d\vec
r'\rho^{\alpha}(\vec r)v(\vec r-\vec r')
\rho^{\alpha}(\vec r')\bigg)}} \nonumber \\
& & =\bigg( \beta \det(v^{-1}) \bigg)^{\frac{1}{2}}
\int[d{\phi}^{\alpha}]\exp{\bigg(-\frac{\beta}{2}\int d\vec
r\phi^{\alpha}(\vec r)v^{-1}\phi^{\alpha}(\vec r)\bigg)}
\exp{\bigg(i\beta\int d\vec r\rho^{\alpha}(\vec
r)\phi^{\alpha}(\vec r)\bigg)} \label{eq4-01}
\end{eqnarray}
and similarly for the core part in (\ref{eq2-07}). The partition function
for the cQGP reads

\begin{equation}
Z=\bigg( \beta \det( v'^{-1}) \bigg)^{\frac{1}{2}({N_c}^2-1)}
\bigg( \det( w^{-1}) \bigg)^{\frac{1}{2}}
\int\prod_{\alpha}^{N_c^2-1}[d{\phi}^{\alpha}][d{\psi}]e^{-S}
\label{eq4-02}
\end{equation}
with the induced action

\begin{eqnarray}
S & = & \sum_{\alpha}^{N_c^2-1}\frac{\beta}{2}\int d\vec
r\phi^{\alpha}(\vec r)v^{-1}\phi^{\alpha}(\vec r)+\frac{1}{2}\int
d\vec r\psi(\vec r)w^{-1}\psi(\vec r) \nonumber \\
& - & \sum_{a}\int dQ_{a}d\vec r n_{a} e^{i\beta
\sum_{\alpha}^{N_c^2-1} Q_{a}^{\alpha}{\phi}^{\alpha}(\vec
r)+i\psi(\vec r) + \frac{1}{2}w_0 } \label{eq4-03}
\end{eqnarray}
Here $w_0$ is the divergent self-energy. For simplicity
the colored particles are point-like throughout.

If we introduce the screened Coulomb potential

\begin{equation}
v_{DH}(\vec r)=\beta \frac{g^2}{4\pi r}e^{-\kappa_a r}
\label{eq4-04}
\end{equation}
with the squared Debye wave number,

\begin{equation}
\kappa_a^2=\frac{g^2}{N_{c}^2-1}\sum_{a}n_{a} \beta
\sum_{\alpha}^{N_c^2-1}{Q_{a}^{\alpha}}^2 \label{eq4-05}
\end{equation}
then the induced action (\ref{eq4-03}) can be split into
a screened part $S_0$ (quadratic in the fields) and an
interaction part $S_I$ (rest). Specifically,

\begin{eqnarray}
\lefteqn{ S=\sum_{\alpha}^{N_c^2-1} \int d\vec r \bigg(
-\frac{1}{N_c^2-1} \sum_{a}n_a + \frac{\beta}{2}
\phi^{\alpha}(\vec r)( v^{-1} + \frac{\kappa_a^2}{g^2})
\phi^{\alpha}(\vec r) \bigg) + \frac{1}{2} \int d\vec r
\psi(\vec r)w^{-1}\psi(\vec r)} \nonumber \\
& & - \sum_{a} \int dQ_{a}d\vec r n_a \bigg( e^{i\beta
\sum_{\alpha} Q_{a}^{\alpha}{\phi}^{\alpha}(\vec r)+i \psi(\vec
r)+ \frac{1}{2}w_0}-1+ \frac{1}{2}({\beta}^2 \sum_{\alpha,
\alpha'} Q_{a}^{\alpha}
Q_{a}^{\alpha'}{\phi}^{\alpha} {\phi}^{\alpha'}) \bigg) \nonumber \\
& & = S_0 + S_{I} \label{eq4-06}
\end{eqnarray}
where we used the color normalization $\int dQ =1$ and the
color averaged squared Debye wave number,

\begin{eqnarray}
\lefteqn{\kappa_a^2\equiv\int dQ_{a} \kappa_a^2=\frac{g^2}{N_{c}^2-1}
\int dQ_{a}\sum_{a}n_{a} \beta \sum_{\alpha}^{N_c^2-1}
{Q_{a}^{\alpha}}^2 } \nonumber \\
& & = \frac{g^2}{N_{c}^2-1} \int dQ_{a}\sum_{a}n_{a} \beta
(N_{c}^2-1) C_{2a} = g^2\sum_{a}n_a \beta C_{2a} \label{eq4-07}
\end{eqnarray}
Here $C_2$ is the quadratic Casimir $(q_2)$ divided by
$(N_c^2-1)$,

\begin{equation}
\int dQ_{a} Q_{a}^{\alpha}Q_{a}^{\alpha'}=C_{2a}
{\delta}^{\alpha\alpha'}=
\frac{q_{2a}}{N_c^2-1}{\delta}^{\alpha\alpha'} \label{eq4-08}
\end{equation}

\subsection{One-Loop}

The screened Debye-Huckel partition function follows by
setting $S_I=0$ in the induced action. The corresponding
partition function is then

\begin{equation}
Z_0= \exp{\bigg( V\sum_{a}n_{a} \bigg)}\bigg(
\det{(1+\frac{1}{-{\vec{\nabla}^2}}\kappa_a^2)}\bigg)^{-\frac{1}{2}({N_c}^2-1)}
\label{eq4-09}
\end{equation}
The argument of the determinant is the inverse screened Green's
function

\begin{equation}
\frac{1}{g^2}\bigg(-{\vec {\nabla}^2}+\kappa_a^2\bigg)G(\vec
r-\vec r')=\delta(\vec r-\vec r') \label{eq4-10}
\end{equation}
which is

\begin{equation}
G(\vec r-\vec r')=g^2\int \frac{d\vec k}{(2\pi)^3} \frac{e^{i\vec
k \cdot (\vec r-\vec r')}}{{\vec k}^2+\kappa_a^2} \label{eq4-11}
\end{equation}
The apparent singularity for coincidental arguments can be
handled by dimensional regularization~\cite{brown&yaffe},

\begin{equation}
\lim_{n\rightarrow 3}G_n(\vec 0)=G(\vec
0)=-\kappa_a\frac{g^2}{4\pi} \label{eq4-12}
\end{equation}
with $n$ being the spatial dimension.  The determinant in (\ref{eq4-09})
can be calculated by standard methods. The identity
$\delta \ln \det X = Tr(X^{-1}\delta X)$ yields

\begin{equation}
\ln \det{(1+\frac{1}{-{\vec {\nabla}^2}}\kappa_a^2)}
=\frac{2}{n}G_n(\vec 0)\frac{\kappa_a^2}{g^2} V \label{eq4-13}
\end{equation}
so that

\begin{equation}
Z_0=\exp{\bigg( V( \sum_{a}n_{a} -
\frac{{N_c}^2-1}{n}\frac{\kappa_a^2}{g^2} G_n(\vec 0)) \bigg)}
\label{eq4-14}
\end{equation}
Using (\ref{eq4-07}) and (\ref{eq4-12}), we obtain the screened
one-loop result as

\begin{equation}
Z_0=\exp{\bigg( V \sum_{a}n_{a} (1+(N_c^2-1)g^2 \beta C_{2a}
\frac{\kappa_a}{12\pi}) \bigg)}=\exp{\bigg( V \sum_{a}n_{a} (1+g^2
\beta q_{2a} \frac{\kappa_a}{12\pi}) \bigg)} \label{eq4-15}
\end{equation}

\subsection{Two-Loop}

\begin{figure}[!h]
\begin{center}
\includegraphics[width=0.20\textwidth]{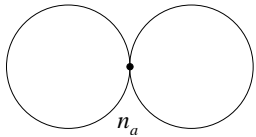}
\end{center}
\caption{Two loop contribution from one particle (\ref{eq4-17})}
\label{2loop1p}
\end{figure}

Higher order loop corrections follow from

\begin{equation}
Z = Z_0 \prod_{\alpha}^{N_c^2-1} \exp{ \bigg( \frac{1}{2\beta}
\int d \vec r_1 d\vec r_2\frac{\delta}{\delta
{\phi}^{\alpha}}G(\vec r_1-\vec r_2)\frac{\delta}{\delta
{\phi}^{\alpha}}\bigg)} \exp{ (-S_I({\phi}^{\alpha}))} {\Bigg
\vert}_{{\phi}^{\alpha}=0} \label{eq4-16}
\end{equation}
by insering higher $G$'s through $S_I$ as recently discussed
in~\cite{brown&yaffe} for the Abelian case. The non-Abelian is
noteworthy in many respects as we note below. A typical two-loop
contribution is shown in Fig.~\ref{2loop1p}. Its contribution is

\begin{figure}[!h]
\begin{center}
\includegraphics[width=0.25\textwidth]{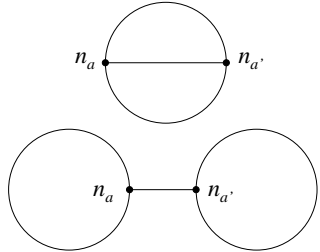}
\end{center}
\caption{Two loop contribution from two particles (\ref{eq4-19})}
\label{2loop2p}
\end{figure}

\begin{equation}
\frac{1}{2^3} \beta^2 (N_c^2-1)^2\sum_{a} n_a {C_{2a}}^2 \int
d\vec rG^2(\vec 0) \label{eq4-17}
\end{equation}
The color factors follow from the identity (\ref{eqc07})
discussed in Appendix B, i.e.

\begin{equation}
\int dQ Q^{\alpha}Q^{\beta}Q^{\gamma}Q^{\delta} = A ( \sum_{n}
d^{\alpha\beta n}d^{\gamma\delta n}+ \sum_{n}d^{\alpha\gamma
n}d^{\beta\delta n}+ \sum_{n}d^{\alpha\delta n}d^{\beta\gamma n}
)+B (\delta^{\alpha\beta}\delta^{\gamma\delta}+
\delta^{\alpha\gamma}\delta^{\beta\delta}+ \delta^{\alpha\delta}
\delta^{\beta\gamma}) \label{eq4-18}
\end{equation}
with $A=0$, $B=1/15 q_2^2$ for SU$(2)$ and $A+3B=3/80
q_2^2$ for SU$(3)$. Although SU(2) and SU(3) involve
considerably different integration measures, the overall
result for Fig.~\ref{2loop1p} is the same. Additional contributions
to the colored partition function are shown in Fig.~\ref{2loop2p},
which contribute

\begin{eqnarray}
\lefteqn{ \quad -\frac{\beta^3}{2\cdot3!} \frac{1}{4} q_3 \int
d\vec r d\vec r'G^3(\vec r-\vec r')\sum_{a}n_{a} \sum_{a'}n_{a'}} \nonumber \\
& & -\frac{\beta^3}{2\cdot4} \int d\vec r d\vec r'G(\vec r-\vec
r')G^2(\vec 0)\sum_{a}n_{a}\sum_{a'}n_{a'}\frac{1}{4^2}
\sum_{\alpha,\beta,\gamma}d^{\alpha\alpha\beta}d^{\beta\gamma\gamma}
\label{eq4-19}
\end{eqnarray}
Here $q_3$ is the cubic Casimir following from the identity

\begin{eqnarray}
&&\sum_{\alpha,\beta,\gamma}^{N_c^2-1}d_{\alpha\beta\gamma}
Q_{a}^{\alpha}Q_{a}^{\beta}Q_{a}^{\gamma}=q_3=\frac{1}{4}\sum_{\alpha,
\beta,\gamma}^{N_c^2-1}d_{\alpha\beta\gamma}d^{\alpha\beta\gamma}
= \int dQ_{a} q_3 = \sum_{\alpha,\beta,\gamma}^{N_c^2-1} \int
dQ_{a}d_{\alpha\beta\gamma}Q_{a}^{\alpha}Q_{a}^{\beta}Q_{a}^{\gamma}
\nonumber \\
&&\int dQ_{a}Q_{a}^{\alpha} Q_{a}^{\beta}Q_{a}^{\gamma} =\frac{1}{4}
d^{\alpha\beta\gamma} \label{eq4-20}
\end{eqnarray}
as detailed in Appendix B. The cubic Casimir exists only for $N_c>2$
and vanishes identically for SU(2). This is clear from the fact that
the contributions in Fig.~\ref{2loop2p} involve 3 colored vertices.
Also note that only the irreducible graph of Fig.~\ref{2loop2p}
contributtes to (\ref{eq4-19}) since the reducible graph averages
to zero by color integration through the identity

\begin{equation}
\sum_{\alpha,\beta,\gamma} d^{\alpha\alpha\beta}
d^{\beta\gamma\gamma}=0 \label{eq4-21}
\end{equation}

Unlike the Abelian case discussed in~\cite{brown&yaffe}, where tadpoles
and disconnected contributions abund, the non-Abelian case has none of
these thanks to the color integrations. Also, the effects of the hard
core at two loop can easily be recalled by noting that in the 2-particle
channel under consideration the distance of minimum encounter is $\sigma$.
So the radial integrations should be limited to $\sigma<r<\infty$ to
account for the hard core. With this in mind, the 2-loop contribution
to the partition function reads

\begin{equation}
\frac{\ln{Z}}{V}=\frac{\ln{Z_0}}{V}+\frac{1}{2^3}\beta^2\sum_{a}
n_a{q_{2a}}^2 G^2(\vec 0)-\sum_{a,a'}n_{a}n_{a'}\Big(\frac{1}
{2\cdot3!}\beta^3\frac{1}{4}q_3\int_{\sigma}^{\infty}d\vec r
G^3(\vec r) \Big) \label{eq4-22}
\end{equation}
where the last term is only present for SU(3). Using (\ref{eq4-15})
and (\ref{eq4-12}) we get

\begin{eqnarray}
\lefteqn{ \frac{\ln{Z}}{V}= (\sum_{a}n_a) + \frac{1}{3}
\sqrt{4\pi} (\frac{g^2}{4\pi})^{\frac{3}{2}} (N_c^2-1)
(\sum_{a}n_a\beta C_{2a})^{\frac{3}{2}} } \nonumber \\
& & + \frac{1}{2^3}4\pi(\frac{g^2}{4\pi})^3(N_c^2-1)^2\Big(
\sum_{a}n_a(\beta C_{2a})^2 \Big)
(\sum_{a}n_{a'}\beta C_{2a}) \nonumber \\
& & - \frac{1}{2 \cdot3!}\beta^3 (\frac{g^2}{4\pi})^3 4\pi
\frac{q_3}{4} (\sum_{a}n_a)^2 E_1(3\kappa_a\sigma)\label{eq4-23}
\end{eqnarray}
Here $E_1(x)$ is the exponential integral$(
-Ei(-x)=E_1(x)=\int_{x}^{\infty}\frac{e^{-t}}{t}dt$), which is
logarithmically divergent at short distance. It is made finite by
the hard core potential.

\subsection{Three Loop}

A partial three loop analysis will be carried out in this section.
There are in total 8 diagrammatic contributions at three loop
that can be organized in terms of the particle density:
1 (one-particle density); 3 (two-particle density); 4 (three- and
four-particle density). They will be considered sequentially.

\begin{figure}[!h]
\begin{center}
\includegraphics[width=0.15\textwidth]{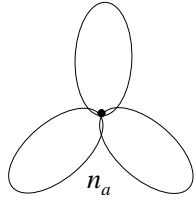}
\end{center}
\caption{Three loop contribution from one particle (\ref{eq4-24})}
\label{3loop1p}
\end{figure}

The three loop contribution stemming from the
one-particle density is shown in  Fi.~\ref{3loop1p}. It is

\begin{equation}
-\frac{1}{3!} \frac{1}{2^3}\beta^3 \sum_{a} n_a \int d\vec r
G^3(\vec 0) {q_{2a}}^3 \label{eq4-24}
\end{equation}
The vertex involves 6 color charges which are integrated with
the help of the identity (see Appendix B)

\begin{eqnarray}
\lefteqn{\int dQ
Q^{\alpha}Q^{\beta}Q^{\gamma}Q^{\delta}Q^{\epsilon}Q^{\zeta} = A
\Big( d^{\alpha\beta\gamma}d^{\delta\epsilon\zeta}
+d^{\alpha\beta\delta}d^{\gamma\epsilon\zeta}
+d^{\alpha\beta\epsilon}d^{\gamma\delta\zeta}
+d^{\alpha\beta\zeta}d^{\gamma\delta\epsilon} } \nonumber \\
& &+d^{\alpha\gamma\delta}d^{\beta\epsilon\zeta}
+d^{\alpha\gamma\epsilon}d^{\beta\delta\zeta}
+d^{\alpha\gamma\zeta}d^{\beta\delta\epsilon}
+d^{\alpha\delta\epsilon}d^{\beta\gamma\zeta}
+d^{\alpha\delta\zeta}d^{\beta\gamma\epsilon}
+d^{\alpha\epsilon\zeta}d^{\beta\gamma\delta} \Big) \nonumber \\
& &+ B \Big(
\delta^{\alpha\beta}\delta^{\gamma\delta}\delta^{\epsilon\zeta} +
\delta^{\alpha\beta}\delta^{\gamma\epsilon}\delta^{\delta\zeta} +
\delta^{\alpha\beta}\delta^{\gamma\zeta}\delta^{\delta\epsilon} +
\delta^{\alpha\gamma}\delta^{\beta\delta}\delta^{\epsilon\zeta} +
\delta^{\alpha\gamma}\delta^{\beta\epsilon}\delta^{\delta\zeta}
\nonumber \\
& &+
\delta^{\alpha\gamma}\delta^{\beta\zeta}\delta^{\delta\epsilon} +
\delta^{\alpha\delta}\delta^{\beta\gamma}\delta^{\epsilon\zeta} +
\delta^{\alpha\delta}\delta^{\beta\epsilon}\delta^{\gamma\zeta} +
\delta^{\alpha\delta}\delta^{\beta\zeta}\delta^{\gamma\epsilon} +
\delta^{\alpha\epsilon}\delta^{\beta\gamma}\delta^{\delta\zeta}
\nonumber \\
& &+
\delta^{\alpha\epsilon}\delta^{\beta\delta}\delta^{\gamma\zeta} +
\delta^{\alpha\epsilon}\delta^{\beta\zeta}\delta^{\gamma\delta} +
\delta^{\alpha\zeta}\delta^{\beta\gamma}\delta^{\delta\epsilon} +
\delta^{\alpha\zeta}\delta^{\beta\delta}\delta^{\gamma\epsilon} +
\delta^{\alpha\zeta}\delta^{\beta\epsilon}\delta^{\gamma\delta}
\Big) \label{eq4-25}
\end{eqnarray}
with $A=0$, $B=\frac{1}{105}q_2^3$ for SU$(2)$ and $A=
-\frac{9}{8!}q_2^3+\frac{27}{2}\frac{1}{7!}q_3^2$, $B=
\frac{85}{2}\frac{1}{8!}q_2^3-\frac{6}{8!}q_3^2$ for SU$(3)$.
Note that this contribution is similar to one-loop for both
SU(2) and SU(3) despite the differences in the contributions
and the color averaging.

\begin{figure}[!h]
\begin{center}
\includegraphics[width=0.27\textwidth]{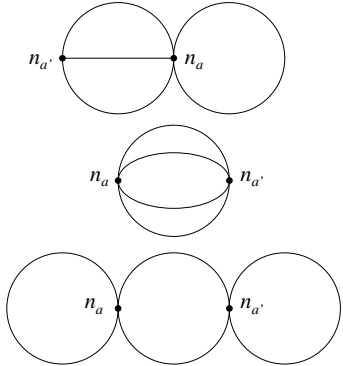}
\end{center}
\caption{Three loop contribution from two particles
(\ref{eq4-26})} \label{3loop2p}
\end{figure}

The three loop contribution stemming from the two-particle density
are shown in Fig~\ref{3loop2p}. The top diagram vanishes for SU(2)
as an odd number of color charges are brought to a single point
which vanish by color averaging. Their contribution is

\begin{eqnarray}
\lefteqn{ \quad \frac{1}{2} \frac{1}{3!} \beta^4 \sum_{a} n_a
q_{2a} \sum_{a'} n_{a'} \frac{q_3}{4} \int d\vec r d\vec r'
G^3(\vec r-\vec r')G(\vec 0) } \nonumber \\
& & + \frac{1}{2^3} \frac{1}{{N_c}^2-1} \beta^4 \sum_{a} n_a
{q_{2a}}^2 \sum_{a'} n_{a'} {q_{2a'}}^2 \int d\vec r d\vec r'
G^2(\vec r'-\vec r)G^2(\vec 0) \nonumber \\
& & + \frac{3}{({N_c}^2-1)({N_c}^2+1)} \frac{1}{4!}\beta^4
\sum_{a} n_a {q_{2a}}^2 \sum_{a'} n_{a'} {q_{2a'}}^2
\int d\vec r d\vec r' G^4(\vec r'-\vec r) \nonumber \\
\label{eq4-26}
\end{eqnarray}

\begin{figure}[!h]
\begin{center}
\includegraphics[width=0.40\textwidth]{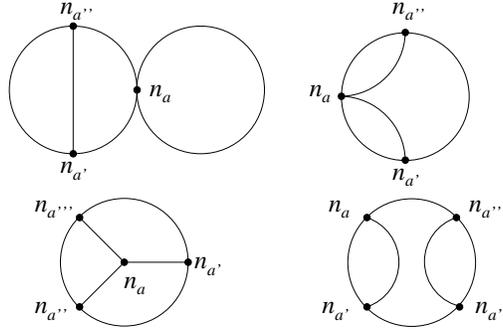}
\end{center}
\caption{Three loop contribution from three and four particle
interactions} \label{3loop34p}
\end{figure}

The three loop contribution stemming from three- and four-particle
density is shown in Fig.~\ref{3loop34p}. These contributions will
not be quoted here. They only contribute for $N_c>2$. They can be
shown to arise from color magnetism, thus subleading in the electric
cQGP under considerations.

\subsection{Higher Loops}

In so far, the color averaging at one- two and three-loops have led
to simple powers of Casimirs. This observation does not carry simply
at higher orders as more complex combination of Casimirs appear.
Indeed, consider the five loop contribution shown in Fig.~\ref{3loop34p}.
Its contribution can be found explicitly as

\begin{figure}[!h]
\begin{center}
\includegraphics[width=0.17\textwidth]{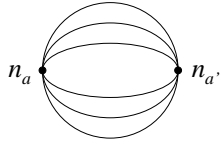}
\end{center}
\caption{A typical five-loop contribution. See text.} \label{5loop2p}
\end{figure}

\begin{equation}
\left\{ \begin{array}{ll}\frac{1}{2} \frac{1}{7!} \beta^6 \sum_{a}
n_a {q_{2a}}^3 \sum_{a'} n_{a'} {q_{2a'}}^3 \int d\vec r d\vec r'
G^6(\vec r'-\vec r) & SU(2) \nonumber \\
\frac{1}{2} \frac{1}{7!} \beta^6 \sum_{a} n_a \sum_{a'} n_{a'}
\bigg( \frac{3}{2^4}{q_3}^2{q_3}^2 - \frac{1}{2^6} {q_3}^2
({q_{2a}}^3+ {q_{2a'}}^3) +\frac{85}{2^8} {q_{2a}}^3{q_{2a'}}^3
\bigg) & \nonumber  \\
\times \int d\vec r d\vec r' G^6(\vec r'-\vec r) & SU(3)
\end{array} \right.
\label{eq4-27}
\end{equation}
The color averaging leads a more complex combination of Casimirs
for SU(3) in comparison to SU(2). The cubic Casimir for SU(3) is
absent for SU(2). The way the color averaging occurs is by noting
that each vertex involves 6 lines in Fig.~\ref{5loop2p}. In SU(2)
the averaging involves only the quadratic Casimir, and the result
with 3 quadratic Casimirs follow. This observation explains why
the SU(2) color averaging in (\ref{eq3-26}) (low density expansion)
can not be extended to SU(3) in closed form. The culpright is the
occurence of the third Casimir.

\subsection{Result}

Now we combine the one, two and (partial) three loop results for
the grand partition function, namely (\ref{eq4-23}), (\ref{eq4-24})
and (\ref{eq4-26}) and use (\ref{eq4-12}), to get

\begin{eqnarray}
\lefteqn{ \frac{\ln{Z}}{V}= (\sum_{a}n_a) + \frac{1}{3}
\sqrt{4\pi} (\frac{g^2}{4\pi})^{\frac{3}{2}} (N_c^2-1)
(\sum_{a}n_a\beta C_{2a})^{\frac{3}{2}}} \nonumber \\
& & + \frac{1}{2^3}4\pi(\frac{g^2}{4\pi})^3(N_c^2-1)^2\Big(
\sum_{a}n_a(\beta C_{2a})^2 \Big)
(\sum_{a}n_{a}\beta C_{2a}) \nonumber \\
& & + \frac{1}{3!}\frac{1}{2^3}(4\pi)^{\frac{3}{2}}
(\frac{g^2}{4\pi})^{\frac{9}{2}} (N_c^2-1)^3
\Big(\sum_{a}n_a(\beta C_{2a})^3 \Big) (\sum_{a}n_{a}\beta
C_{2a})^{\frac{3}{2}} \nonumber \\
& & + \frac{1}{16} ( 4\pi )^{\frac{3}{2}} \Big( \frac{g^2}{4\pi}
\Big)^{\frac{9}{2}} ({N_c}^2-1)^3 \Big( \sum_{a} n_a (\beta
C_{2a})^2 \Big)^2 \Big( \sum_{a} n_{a} (\beta C_{2a})
\Big)^{\frac{1}{2}} e^{-2\kappa_a \sigma} \nonumber \\
& & - \frac{3}{{N_c}^2+1} \frac{1}{4!} 4\pi \Big( \frac{g^2}{4\pi}
\Big)^4 ({N_c}^2-1)^3 \Big( \sum_{a} n_a (\beta C_{2a})^2 \Big)^2
\frac{1}{\sigma}\Big(e^{-4\kappa_a\sigma}+4\kappa_a\sigma
E_1(4\kappa_a\sigma)\Big) \label{eq4-28}
\end{eqnarray}
for both SU(2) and SU(3), with in addition for SU(3) alone

\begin{eqnarray}
& & - \frac{1}{2} \frac{1}{3!} (\frac{g^2}{4\pi})^3 4\pi
(\sum_{a}n_a)^2 \beta^3 \frac{q_3}{4}E_1(3\kappa_a\sigma) \nonumber \\
& & - \frac{1}{2} \frac{1}{3!} ( 4\pi)^{\frac{3}{2}} \Big(
\frac{g^2}{4\pi} \Big)^{\frac{9}{2}} ({N_c}^2-1) \Big( \sum_{a}
n_a (\beta C_{2a}) \Big)^{\frac{3}{2}}  \sum_{a} n_{a}  \beta^3
\frac{q_3}{4} E_1(3\kappa_a\sigma) \nonumber \\
\label{eq4-29}
\end{eqnarray}
To proceed further, we set all classical fugacities for the
three species $a=1,2,3$  (quark, antiquark, gluon) to be the
same, $n_a=n_{a'}=\lambda$. We also rescale all dimensions
to the core size $\sigma$ to be reinstated by inspection as
announced. Specifically $\tilde{\lambda}=\sigma^3 \lambda$,
with all radial integrations cutoff by the core. Using
(\ref{eq4-28}) and (\ref{eq4-29}) with
$\varepsilon_{a}=\frac{g^2}{4\pi}\beta C_{2a}/\sigma$,
we have for both SU(2) and SU(3)

\begin{eqnarray}
\lefteqn{ \frac{\sigma^3\ln{Z}_{\tilde{\lambda}}}{V}=
\tilde{\lambda} \sum_{a} + {\tilde{\lambda}}^{\frac{3}{2}}
\frac{1}{3} (4\pi)^{\frac{1}{2}} (N_c^2-1)(\sum_{a}\varepsilon_{a}
)^{\frac{3}{2}} + {\tilde{\lambda}}^{2}\frac{1}{8}4\pi(N_c^2-1)^2
(\sum_{a}{\varepsilon_{a}}^2)(\sum_{a}\varepsilon_{a})} \nonumber
\\& & - {\tilde{\lambda}}^2 \frac{3}{{N_c}^2+1} \frac{1}{4!}  4\pi
(N_c^2-1)^3 ( \sum_{a} {\varepsilon_{a}}^2 )^2 \Big(e^{-4\kappa_a
\sigma}+4\kappa_a\sigma E_1(4\kappa_a\sigma)\Big) \nonumber \\
& & + {\tilde{\lambda}}^{\frac{5}{2}}
\frac{1}{3!}\frac{1}{8}(4\pi)^{\frac{3}{2}} (N_c^2-1)^3
(\sum_{a}{\varepsilon_{a}}^3)
(\sum_{a}\varepsilon_{a})^{\frac{3}{2}} +
{\tilde{\lambda}}^{\frac{5}{2}} \frac{1}{16} (4\pi)^{\frac{3}{2}}
 (N_c^2-1)^3(\sum_{a} {\varepsilon_{a}}^2)^2(\sum_a \varepsilon_{a}
)^{\frac{1}{2}}e^{-2\kappa_a\sigma} \nonumber \\
\label{eq4-30}
\end{eqnarray}
while in addition for SU(3),

\begin{eqnarray}
& &- {\tilde{\lambda}}^{2} \frac{1}{2}\frac{1}{3!} 4\pi \Big(
\frac{g^2}{4\pi} \Big)^3 (\sum_{a})^2 (\frac{\beta^3
q_3}{4\sigma^3})E_1(3\kappa_a\sigma) \nonumber \\
& & -{\tilde{\lambda}}^{\frac{5}{2}}  \frac{1}{2}\frac{1}{3!} (
4\pi)^{\frac{3}{2}} \Big( \frac{g^2}{4\pi} \Big)^3 ({N_c}^2-1)
(\sum_{a}\varepsilon_{a})^{\frac{3}{2}}\sum_{a} (\frac{\beta^3q_3}
{4\sigma^3}) E_1(3\kappa_a\sigma) \nonumber \\
\label{eq4-31}
\end{eqnarray}

We note that the core integrations stemming from $\mid \vec r-\vec
r' \mid < \sigma$ reduce to

\begin{eqnarray}
{\frac{\ln{Z}}{V}}_{\mid \vec r-\vec r' \mid < \sigma} & &=
-\lambda^2\frac{1}{4}(N_c^2-1)
\Big(\sum_{a}\beta C_{2a}\Big)^2\int d\vec r G^2(r) \nonumber \\
& & +\lambda^2\frac{1}{4}(N_c^2-1)^2\Big(\sum_{a}(\beta
C_{2a})^2\Big)\Big(\sum_{a}\beta C_{2a}\Big) \int d\vec r
G^2(r)G(0) \nonumber \\
& & -\lambda^2\frac{1}{8}(N_c^2-1)^3\Big(\sum_{a}(\beta
C_{2a})^3\Big)\Big(\sum_{a}\beta C_{2a}\Big) \int d\vec r
G^2(r)G^2(0) \label{eq4-32}
\end{eqnarray}

where the divergences are lumped in $G(0)$. Our final result
for SU(2) and SU(3) is then

\begin{eqnarray}
 \frac{\sigma^3\ln{Z_{\tilde{\lambda}}}}{V}& &=
\tilde{\lambda} \sum_{a} + {\tilde{\lambda}}^{\frac{3}{2}}
\frac{1}{3} ( 4\pi)^{\frac{1}{2}} (N_c^2-1)
(\sum_{a}\varepsilon_{a})^{\frac{3}{2}}  \nonumber \\
& &-{\tilde{\lambda}}^{\frac{3}{2}}\frac{1}{8}(4\pi)^{\frac{1}{2}}
(N_c^2-1)(\sum_{a}\varepsilon_{a})^{\frac{3}{2}}(1-e^{-2(4\pi)^{
\frac{1}{2}}\tilde{\lambda}^{\frac{1}{2}}(\sum_a\varepsilon_a)^{
\frac{1}{2}}}) \nonumber \\
& &-{\tilde{\lambda}}^2\frac{1}{8}(4\pi)
(N_c^2-1)^2(\sum_{a}\varepsilon_{a})(\sum_{a}{\varepsilon_{a}}^2)
(1-e^{-2(4\pi)^{\frac{1}{2}}\tilde{\lambda}^{\frac{1}{2}}
(\sum_a\varepsilon_a)^{\frac{1}{2}}}) \nonumber \\
& & + {\tilde{\lambda}}^{2} \frac{1}{8}(4\pi)(N_c^2-1)^2 (\sum_{a}
\varepsilon_{a})(\sum_{a}{\varepsilon_{a}}^2)\nonumber \\
& & - {\tilde{\lambda}}^2 \frac{3}{{N_c}^2+1} \frac{1}{4!}(4\pi)
(N_c^2-1)^3 ( \sum_{a} {\varepsilon_{a}}^2)^2 \nonumber \\
& &\times \Big(e^{-4(4\pi)^{\frac{1}{2}}\tilde{\lambda}^{\frac{1}
{2}}(\sum_a\varepsilon_a)^{\frac{1}{2}}}+4(4\pi)^{\frac{1}{2}}
\tilde{\lambda}^{\frac{1}{2}}(\sum_a\varepsilon_a)^{\frac{1}{2}}
E_1(4(4\pi)^{\frac{1}{2}}\tilde{\lambda}^{\frac{1}{2}}(\sum_a
\varepsilon_a)^{\frac{1}{2}})\Big) \nonumber \\
& &-{\tilde{\lambda}}^{\frac{3}{2}}\frac{1}{16}(4\pi)^{\frac{3}
{2}}(N_c^2-1)^3(\sum_{a}\varepsilon_{a})^{\frac{3}{2}}(\sum_{a}
{\varepsilon_{a}}^3)(1-e^{-2(4\pi)^{\frac{1}{2}}\tilde{\lambda}^{\frac{1}{2}}
(\sum_a\varepsilon_a)^{\frac{1}{2}}}) \nonumber \\
& & + {\tilde{\lambda}}^{\frac{5}{2}} \frac{1}{3!}\frac{1}{8}
(4\pi)^{\frac{3}{2}}(N_c^2-1)^3(\sum_{a}\varepsilon_{a})^{\frac{3}{2}}
(\sum_{a}{\varepsilon_{a}}^3) \nonumber \\
& & + {\tilde{\lambda}}^{\frac{5}{2}} \frac{1}{16} (
4\pi)^{\frac{3}{2}} (N_c^2-1)^3( \sum_{a} {\varepsilon_{a}}^2 )^2
(\sum_a \varepsilon_{a})^{\frac{1}{2}}e^{-2(4\pi)^{\frac{1}{2}}
\tilde{\lambda}^{\frac{1}{2}}(\sum_a\varepsilon_a)^{\frac{1}{2}}}
\label{eq4-33}
\end{eqnarray}
with in addition for SU(3)

\begin{eqnarray}
& & -{\tilde{\lambda}}^{2} \frac{1}{2}\frac{1}{3!} 4\pi \Big(
\frac{g^2}{4\pi} \Big)^3 (\sum_{a})^2 (\frac{\beta^3
q_3}{4\sigma^3})E1\Big(3(4\pi)^{\frac{1}{2}}\tilde{\lambda}^{\frac{1}{2}}
(\sum_a\varepsilon_a)^{\frac{1}{2}}\Big) \nonumber \\
& & -{\tilde{\lambda}}^{\frac{5}{2}}  \frac{1}{2}\frac{1}{3!} (
4\pi)^{\frac{3}{2}} \Big( \frac{g^2}{4\pi} \Big)^3 ({N_c}^2-1)
(\sum_{a}\varepsilon_{a})^{\frac{3}{2}}\sum_{a} (\frac{\beta^3q_3}
{4\sigma^3})E1\Big(3(4\pi)^{\frac{1}{2}}\tilde{\lambda}^{\frac{1}{2}}
(\sum_a\varepsilon_a)^{\frac{1}{2}}\Big)\label{eq4-34}
\end{eqnarray}

This our final result for up to three loops ignoring the diagrams
of Fig.~\ref{3loop34p}. The latters yield contributions that are
of order $\tilde{\lambda}^3$ or higher.

\subsection{High Temperature}

The results of the high-temperature expansion can be recovered
from the three loop results~(\ref{eq4-33}). For that we need to go
back to the unintegrated contributions in r-space and expand in
powers of $\beta$ the exponentials. Up to order $\beta^3$ the
result is

\begin{eqnarray}
\frac{\ln{Z}}{V}& &= \lambda \sum_{a} +
\lambda^{\frac{3}{2}}\frac{1}{3}(4\pi)^{
\frac{1}{2}}(N_c^2-1)(\sum_{a}\gamma_{a})^{\frac{3}{2}} \nonumber \\
& & -\lambda^2\frac{1}{4}(N_c^2-1)(\sum_{a}\gamma_a)^2
\int_{r<\sigma} d\vec r \frac{1}{r^2}+\lambda^{\frac{5}{2}}
\frac{1}{2}(4\pi)^{\frac{1}{2}}(N_c^2-1)(\sum_{a}\gamma_a)^{
\frac{5}{2}}\int_{r<\sigma} d\vec r\frac{1}{r} \nonumber \\
& & -\lambda^3\frac{1}{2}(4\pi)(N_c^2-1)(\sum_{a}\gamma_a)^3
\int_{r<\sigma} d\vec r + \lambda^{2} \frac{1}{8} 4\pi
(N_c^2-1)^2(\sum_{a} {\gamma_{a}}^2)(\sum_{a}\gamma_{a})
\nonumber \\
& & -{\lambda}^{2} \frac{1}{2}\frac{1}{3!} \Big(
\frac{g^2}{4\pi}\Big)^3(\sum_{a})^2(\frac{\beta^3q_3}{4})
\int_{\sigma<r<\infty}d\vec r \frac{1}{r^3}e^{-3(4\pi)^{
\frac{1}{2}}\lambda^{\frac{1}{2}}(\sum_a\gamma_a)^{\frac{1}{2}}r}
+\mathcal{O}(\beta^{\frac{7}{2}}) \label{eq4-35}
\end{eqnarray}
where we used $\epsilon_a=\gamma/\sigma$ from (\ref{eq2-10}). For
SU(2) the latter contribution is absent. This result agrees with
(\ref{eq3-34}) to order $\beta^{5/2}$ for
$h_0^{(2)}(r)=-\theta(\sigma -r)$. There is a difference to order
$\beta^3$ which can be traced back to the handling of the core
potential $h_0^{(2)}(r)$ and $w(r)$ in each of the two expansions.
In the loop expansion we used an infinite core with $e^{-w(r)}=1$
for $r>\sigma$. In reality $e^{-w(r)}$ is a function of $r$ much
like $h_0^{(2)}(r)$ when $r>\sigma$. A similar observation was
made for hard sphere liquids in~\cite{raimbault&caillol}.

\section{Free Energies of cQGP}

\renewcommand{\theequation}{VII.\arabic{equation}}
\setcounter{equation}{0}

The free energy follows from the grand partition function which we
have now constructed for both the density (virial) expansion and
the loop (high) temperature expansion. The latter is understood
for a liquid of hard spheres with $h_0^{(2)}(r)=-\theta(\sigma
-r)$ as in~(\ref{eq4-35}) up to order $\beta^{5/2}$. The former is
given in~(\ref{eq3-25}) up to order $\lambda^2$.

Specializing to SU(2) one component plasma($\sum_a\gamma_a
\to\gamma$) in (\ref{eq4-35}) and carrying the $r$-integration
with the hard core potential in mind as noted, we have

\begin{eqnarray}
\frac{\ln{Z}}{V}& &= \lambda +\lambda^{\frac{3}{2}}\frac{1}{3}
(4\pi)^{\frac{1}{2}}(N_c^2-1)\gamma^{\frac{3}{2}}
-\lambda^2\pi(N_c^2-1)\gamma^2\sigma+2\pi^{\frac{3}{2}}
\lambda^{\frac{5}{2}}(N_c^2-1)\gamma^{\frac{5}{2}}\sigma^2 \nonumber \\
& & -\lambda^3\frac{8}{3}\pi^2(N_c^2-1)\gamma^3\sigma^3  +
\lambda^{2} \frac{1}{2}\pi(N_c^2-1)^2\gamma^3
+\mathcal{O}(\beta^{\frac{7}{2}}) \label{eq6-01}
\end{eqnarray}

\noindent The density
$c (=\lambda\frac{\partial}{\partial\lambda}(\frac{\ln{Z}}{V}))$ is

\begin{equation}
c=\lambda+\triangle c=\lambda+\sqrt{\pi}(N_c^2-1)\gamma^{\frac{3}
{2}}\lambda^{\frac{3}{2}}+\mathcal{O}(\beta^2)
\label{eq6-02}\end{equation}
The corresponding shift induced in the chemical potential,
$\beta\mu_c=\beta\mu_{\lambda}+\beta\triangle\mu_c$, due
to the interactions can be extracted from

\begin{equation}
c=\frac{g}{\Lambda^3}e^{\beta\mu_c}=\frac{g}{\Lambda^3}
e^{\beta\mu_{\lambda}+\beta\triangle\mu_c}\simeq\frac{g}{\Lambda^3}
e^{\beta\mu_{\lambda}}(1+\beta\triangle\mu_c)=\lambda+\lambda\beta
\triangle\mu_c=\lambda+\triangle c \label{eq6-03}\end{equation}
with

\begin{equation}
\beta\triangle\mu_c=\frac{\triangle c}{\lambda}=\sqrt{\pi}
(N_c^2-1)\gamma^{\frac{3}{2}}\lambda^{\frac{1}{2}}+\mathcal{O}(\beta^2)\,\,.
\label{eq6-04}
\end{equation}
Using this shift and defining the free energy through the legendre
transform

\begin{eqnarray}
\beta\frac{F(\beta,c)}{V} =-\frac{1}{V} \ln{Z(\beta,
\lambda)}+c\beta\mu_c\end{eqnarray} we obtain for either the loop
expansion or the high temperature expansion

\begin{eqnarray}
\beta\frac{F_{loop}(\beta,c)}{V}& &
=-c+\frac{2\sqrt{\pi}}{3}(N_c^2-1)\lambda^{\frac{3}{2}}
\gamma^{\frac{3}{2}}+\pi(N_c^2-1)\lambda^2\gamma^2\sigma \nonumber \\
& &-2\pi^{\frac{3}{2}}(N_c^2-1)\lambda^{\frac{5}{2}}\gamma^{\frac
{5}{2}}\sigma^2-\frac{\pi}{2}(N_c^2-1)^2\lambda^2\gamma^3+c\beta\mu_{\lambda}
+\mathcal{O}(\beta^{\frac{7}{2}}) \nonumber \\
& &=-\lambda-\frac{2\sqrt{\pi}}{3}
(N_c^2-1)\gamma^{\frac{3}{2}}\lambda^{\frac{3}{2}}-\sqrt{\pi}
\lambda^{\frac{1}{2}}(N_c^2-1)\gamma^{\frac{3}{2}}\bigg(\sqrt{\pi}
(N_c^2-1)\gamma^{\frac{3}{2}}\lambda^{\frac{3}{2}}\bigg) \nonumber \\
& &+\frac{1}{2}\pi(N_c^2-1)^2\gamma^3\lambda^2+\pi(N_c^2-1)
\lambda^2\gamma^2\sigma-2\pi^{\frac{3}{2}}(N_c^2-1)\lambda^{\frac{5}{2}}
\gamma^{\frac{5}{2}}\sigma^2+c\beta\mu_{\lambda}+\mathcal{O}(\beta^{\frac{7}{2}})
\nonumber \\
& &=-\lambda-\frac{2\sqrt{\pi}}{3}
(N_c^2-1)\gamma^{\frac{3}{2}}\lambda^{\frac{3}{2}}-\sqrt{\pi}
\lambda^{\frac{1}{2}}(N_c^2-1)\gamma^{\frac{3}{2}}\bigg(\triangle
c +\mathcal{O}(\beta^2)\bigg)+\frac{1}{2\lambda}(\triangle c)^2\nonumber \\
& &+\pi(N_c^2-1)\lambda^2\gamma^2\sigma-2\pi^{\frac{3}{2}}
(N_c^2-1) \lambda^{\frac{5}{2}}\gamma^{\frac{5}{2}}\sigma^2
+c\beta\mu_{\lambda}+\mathcal{O}(\beta^{\frac{7}{2}})\nonumber \\
& &\simeq-\lambda\Big(1+\frac{\triangle c}{\lambda}\Big)
-\frac{2\sqrt{\pi}}{3}(N_c^2-1)\gamma^{\frac{3}{2}}\lambda^{\frac{3}{2}}
\Big(1+\frac{\triangle c}{\lambda}\Big)^{\frac{3}{2}}+\pi(N_c^2-1)
\lambda^2\Big(1+\frac{\triangle c}{\lambda}\Big)^2\gamma^2\sigma \nonumber \\
& &-2\pi^{\frac{3}{2}}(N_c^2-1)\lambda^{\frac{5}{2}}\Big(1+
\frac{\triangle c}{\lambda}\Big)^{\frac{5}{2}}\gamma^{\frac{5}{2}}
\sigma^2 +c(\beta\mu_{\lambda}+\beta\triangle \mu_c)+\mathcal{O}(\beta^3) \nonumber \\
& &=-c-\frac{2\sqrt{\pi}}{3}(N_c^2-1)\gamma^{\frac{3}{2}}
c^{\frac{3}{2}}+\pi(N_c^2-1)c^2\gamma^2\sigma-2\pi^{\frac{3}{2}}(N_c^2-1)
c^{\frac{5}{2}}\gamma^{\frac{5}{2}}\sigma^2 \nonumber \\
& & +c\beta\mu_c+\mathcal{O}(\beta^3)\,\,. \label{eq6-05}
\end{eqnarray}
From (\ref{eq6-02}), we note that $\triangle c\sim\beta^{\frac{3}{2}}$
so that $\lambda$ can be substituted by the concentration $c$ for terms
that are of order $\beta^2$ or higher to accuracy $\beta^{5/2}$. For the
free energy, the high temperature expansion deviates from the loop-expansion
by such substitutions in higher order.

The free energy in the low density expansion can be extracted from
(\ref{eq3-25}) using a similar transform and substitution. For
SU(2) the procedure of subtitution of $\tilde{\lambda}$ by
$\tilde{c}$ is detailed in~\cite{borisetal2}.  For SU(2)
$D_2=B_2+\frac{3}{8}B_{3/2}^2$ where the B's are Bernoulli's
numbers. For a one species SU(2) plasma $\epsilon=\sqrt{\epsilon_a
\epsilon_{a}}=\epsilon_a$. Recalling that in~\cite{borisetal2} the
normalizations were carried using the Wigner-Seitz radius instead
of the core size with $\delta=\sigma/a_{WS}$, we have
$\epsilon={\Gamma}/{\delta}$. The SU(2) free energy of a
one-species cQGP in the low density expansion is

\begin{eqnarray}
\beta\frac{F_{low}(\beta,c)}{V}&=&-c-2\sqrt{\pi}\gamma^{\frac{3}{2}}
c^{\frac{3}{2}}-3\pi\gamma^3c^2-\frac{9}{4}\pi\gamma^3c^2
\int_{0}^{\frac{3}{\sigma}\gamma}\frac{\sinh t }{t}dt
\nonumber \\
&+&\frac{\pi}{12}\frac{\sigma^4}{\gamma}c^2\bigg(2\sinh({\frac{3}{\sigma}
\gamma})+2\frac{\gamma}{\sigma}\cosh({\frac{3}{\sigma}\gamma})+3\frac{
\gamma^2}{\sigma^2}\sinh({\frac{3}{\sigma}\gamma})+9\frac{\gamma^3}{\sigma^3}
\cosh({\frac{3} {\delta}\gamma})\bigg) \nonumber \\
&+& c\beta\mu_c+\mathcal{O}(c^{\frac{5}{2}}) \,\,.\label{eq6-13}
\end{eqnarray}

\section{Excess Free Energies for cQGP}

\renewcommand{\theequation}{VIII.\arabic{equation}}
\setcounter{equation}{0}

The free energies obtained above can be rewritten in terms of
the plasma constant

\begin{equation}
\Gamma=\frac{g^2}{4\pi}\frac{C_{2}}{T a_{WS}} \label{eq6-14}
\end{equation}
with $k_B=1$ and $a_{WS}$ the Wigner-Seitz radius
satisfying $N/V(4\pi a_{WS}^3/3)=1$. $C_2$ is the quadratic
Casimir, $C_2=q_2/(N_c^2-1)$ and $g$ is the strength of the
coupling. Since the Wigner-Seitz radius is related to the
density, $N/V=c$, it is straightforward to rewrite the
expanded free energies above in terms of $\Gamma$. For instance,
the first two terms in the loop expansion (\ref{eq6-05}) read

\begin{eqnarray}
\frac{F_{loop}(\Gamma)}{NT}&=&-c\frac{V}{N}-c^{\frac{3}{2}}
\frac{V}{N}\frac{2\sqrt{\pi}}{3}(\frac{g^2}{4\pi})^{\frac{3}{2}}
(N_c^2-1)(\beta C_{2})^{\frac{3}{2}} \nonumber \\
&=&-1-c^{\frac{3}{2}}\frac{4\pi}{3}a_{WS}^3\frac{(4\pi)^{\frac{1}
{2}}}{3} (\frac{g^2}{4\pi})^{\frac{3}{2}}(N_c^2-1)(\beta
C_{2})^{\frac{3}{2}} \nonumber \\
&=&-1-\frac{1}{\sqrt{3}}c^{\frac{3}{2}} \Big( \frac{4\pi
a_{WS}^3}{3}\Big)^{\frac{3}{2}} (N_c^2-1) (\frac{g^2}{4\pi}\beta
\frac{C_{2}}{a_{WS}})^{\frac{3}{2}} \nonumber \\
&=& -1-\frac{1}{\sqrt{3}} (N_c^2-1) \Gamma^{\frac{3}{2}}
\label{eq6-15}
\end{eqnarray}
If we define the excess free energy $F_{ex}$ as $F(\Gamma)=
F(0)+F_{ex}(\Gamma)$, we get the excess free energies for
(\ref{eq6-05}) and (\ref{eq6-13}) as

\begin{eqnarray}
\frac{F_{loop,ex}}{NT}&=&-\frac{1}{\sqrt{3}}(N_c^2-1)\Gamma^{\frac{3}{2}}
+\frac{3}{4}\delta(N_c^2-1)\Gamma^2-3\sqrt{3}\delta^2(N_c^2-1)
\Gamma^{\frac{5}{2}}+\mathcal{O}(\Gamma^3) \nonumber \\
\frac{F_{low,ex}}{NT}&=&-\sqrt{3}\Gamma^{\frac{3}{2}}
-\frac{9}{4}\Gamma^3-\frac{27}{16}\Gamma^3
\int_{0}^{\frac{3}{\delta}\Gamma}\frac{\sinh t }{t}dt
\nonumber \\
&+&\frac{1}{16}\frac{\delta^4}{\Gamma}\bigg(2\sinh({\frac{3}{\delta}
\Gamma})+2\frac{\Gamma}{\delta}\cosh({\frac{3}{\delta}\Gamma})+3\frac{
\Gamma^2}{\delta^2}\sinh({\frac{3}{\delta}\Gamma})+9\frac{\Gamma^3}{\delta^3}
\cosh({\frac{3} {\delta}\gamma})\bigg)  \label{eq6-16}
\end{eqnarray}

To compare these expanded results with the full molecular dynamics
simulations in~\cite{borisetal}, we may also write

\begin{eqnarray}
\frac{F_{mol,ex}}{NT}&=&-4.9\int_{+0}^{\Gamma}\frac{d\Gamma'}{\Gamma'}
-2\Gamma+3.2\cdot4\Gamma^{\frac{1}{4}}-2.2\cdot4\Gamma^{-\frac{1}{4}}
+\frac{F_{mol, ex}(+0)}{NT} \label{eq6-17}
\end{eqnarray}
with $A=2/3(2/(\frac{\pi}{2}N_c(N_c-1))^2)^{1/3}$. $F_{mol,ex}$
was obtained from the potential energy (36) in~\cite{borisetal}
through the following relation

\begin{equation}
\frac{F_{ex}(\Gamma)}{NT}=\int_{+0}^{\Gamma}\frac{U_{ex}}{NT}
\frac{d\Gamma'}{\Gamma'}+\frac{F_{ex}(+0)}{NT} \,\,.\label{eq6-18}
\end{equation}
For completeness we also quote the excess energy in the Debye-Huckel
limit~\cite{borisetal2},

\begin{eqnarray}
\frac{F_{DH,ex}}{NT}&=&-(N_c^2-1)\frac{3d}{2}\bigg(
(A\Gamma)^{\frac{3}{2}}\tan^{-1}\Big(\frac{1}{\sqrt{A\Gamma}}\Big)
-\frac{1}{2}\Big(\ln(1+A\Gamma)-A\Gamma\Big)\bigg)\,\,.
\label{eq6-19}
\end{eqnarray}

\begin{figure}[!h]
\begin{center}
\includegraphics[width=0.49\textwidth]{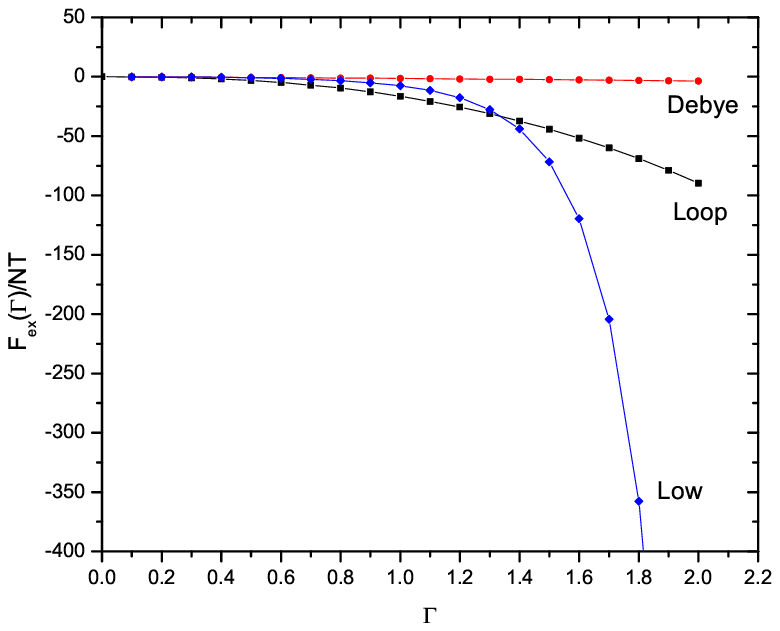}
\includegraphics[width=0.50\textwidth]{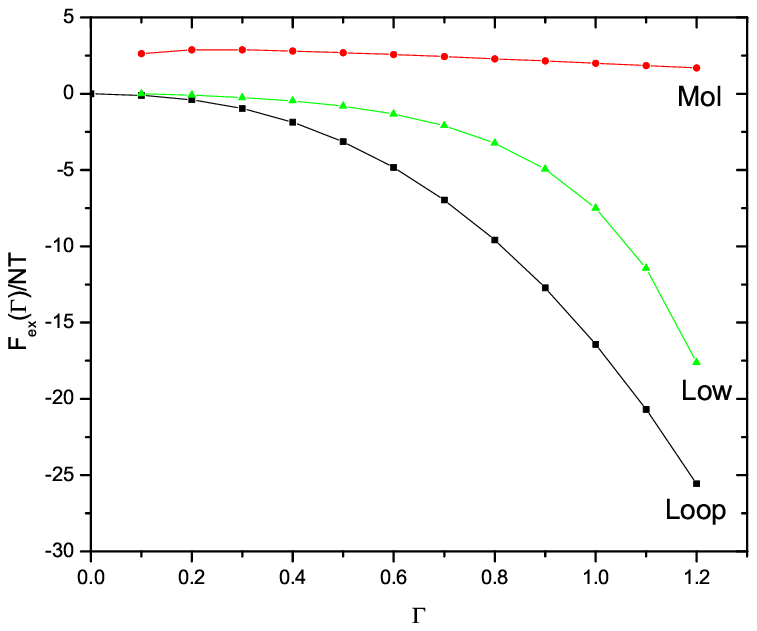}
\end{center}
\caption{Excess free energies vs Debye-Huckel (a) and molecular
dynamics (b). See text.}
\label{free_energy}
\end{figure}

In Fig.~\ref{free_energy}a we show the behavior of the excess free
energies for the low density and loop expansions in comparison to
the Debye-Huckel result. By about $\Gamma\approx 1.5$ the three
expansions deviate substantially and are no longer reliable for
the free energy. The deviation occurs earlier in the energy
density as we discussed in~\cite{cho&zahed}. In
Fig.~\ref{free_energy}b we compare the three expansions to the
molecular dynamics simulations in~\cite{borisetal} for
$0.2<\Gamma<1.2$. We note that the ${\rm ln}(+0)$ contribution in
(\ref{eq6-17}) combines with $F(+0)$ to zero(Infrared
renormalization). The molecular dynamics simulations use a finite
3-box in space which at weak coupling does not accomodate the
Debye cloud which lowers (attractive) the free energy. Thus the
discrepancy with the Debye-Huckel limit which is exact for a
classical plasma. Simulations with larger 3-boxes are possible but
numerically time consuming. At strong coupling, the molecular
simulations readily accomodate the short range liquid and/or
crystal correlations with the current finite 3-boxes. They are
more reliable at large $\Gamma$.

\section{Conclusions}

We have explicitly analyzed the grand partition function of a cQGP
defined by colored spheres with a hard core potential and a long
range Coulomb field, borrowing on methods used for classical
liquids~\cite{caillol&raimbault}. The partition function was
worked out explicitly for both SU(2) and SU(3) at low densities
through a cumulant expansion, and at high temperature through a
loop expansion. Both expansions are shown to be valid for low
plasma coupling $\Gamma=V/K\approx 1$. This is the regime where
the quantum QGP is expected to be dilute. Current interest in
the quantum QGP is in the range of $\Gamma\approx 3$ were the
Coulomb interactions are much stronger and the phase liquid-like.
In the follow up paper~\cite{cho&zahed} we construct an equation
of state that interpolates smoothly between the expanded results
of this paper at low $\Gamma$ and the strongly coupled molecular
dynamics results at large $\Gamma$~\cite{borisetal}.

\section{Acknowledgments}
This work was supported in part by US-DOE grants DE-FG02-88ER40388 and
DE-FG03-97ER4014.

\appendix

\section{Color charges}

\renewcommand{\theequation}{A.\arabic{equation}}
\setcounter{equation}{0}

The set of classical color charges form a Darboux's set

\begin{equation}
\vec{\phi}=(\phi_1,\cdots, \phi_{\frac{1}{2}N_c(N_c-1)}) \quad
\vec{\pi}=(\pi_1,\cdots, \pi_{\frac{1}{2}N_c(N_c-1)})
\label{eqb01}
\end{equation}
with canonical Poisson brackets

\begin{equation}
\{ \phi_{\alpha}, \pi_{\beta} \} = \delta_{\alpha\beta} \quad
\{Q^{\alpha}, Q^{\beta} \} = f^{\alpha\beta\gamma}Q^{\gamma}
\label{eqb02}
\end{equation}
where $N_c$ is the number of colors and $f^{\alpha\beta\gamma}$ is
the structure constant of SU$(N_c)$. The explicit SU(2) and SU(3)
representations of these sets can be found
in~\cite{johnson,litim&manuel}. In this Appendix, we quote some
formulas for completeness.

The Darboux's set for SU(2) is~\cite{litim&manuel}

\begin{equation}
Q^1=\cos\phi_1 \sqrt{J^2-\pi_1^2}, \quad  Q^2=\sin\phi_1
\sqrt{J^2-\pi_1^2}, \quad Q^3=\pi \label{eqb03}
\end{equation}

\noindent where $J^2$ is used to represent the quadratic Casimir
with $q_2= \sum_{\alpha}^{N_c^2-1}{Q^{\alpha}Q^{\alpha}}$. The
phase space SU(2) measure used in the text is

\begin{equation}
dQ=c_R d\pi_1 d\phi_1 J dJ \delta(J^2-q_2) \label{eqb04}
\end{equation}
The representation dependent constant $c_R$ is chosen so that $\int dQ=1$.

The Darboux's set for SU(3) is more involved. Following~\cite{litim&manuel}
we define the canonical set as $(\pi_1,\pi_2,\pi_3,\phi_1,\phi_2,\phi_3,J_1,J_2)$
in terms of which the 8 color charges read

\begin{eqnarray}
& & Q^1=\cos\phi_1\pi_{+}\pi_{-} \quad
Q^2=\sin\phi_1\pi_{+}\pi_{-}
\quad Q^3=\pi_1 \quad Q^8=\pi_2 \nonumber \\
& & Q^4=C_{++}\pi_{+}A+C_{+-}\pi_{-}B \quad\quad
Q^5=S_{++}\pi_{+}A+S_{+-}\pi_{-}B \nonumber \\
& & Q^6=C_{--}\pi_{-}A-C_{--}\pi_{+}B \quad\quad
Q^7=S_{--}\pi_{-}A-S_{--}\pi_{+}B \nonumber \\
\label{eqb05}
\end{eqnarray}
with

\begin{eqnarray}
\pi_{+}=\sqrt{\pi_3+\pi_1}, \quad \quad
C_{\pm\pm}=\cos\Big[\frac{1}{2}(\pm\phi_1+\sqrt{3}\phi_2\pm\phi_3)\Big]
\nonumber \\
\pi_{-}=\sqrt{\pi_3-\pi_1}, \quad \quad
S_{\pm\pm}=\sin\Big[\frac{1}{2}(\pm\phi_1+\sqrt{3}\phi_2\pm\phi_3)\Big]
\label{eqb06}
\end{eqnarray}
and

\begin{eqnarray}
A=\frac{1}{2\pi_3}\sqrt{(\frac{J_1-J_2}{3}+\pi_3+\frac{\pi_2}{\sqrt{3}})
(\frac{J_1+2J_2}{3}+\pi_3+\frac{\pi_2}{\sqrt{3}})(\frac{2J_1+J_2}{3}-\pi_3-\frac{\pi_2}{\sqrt{3}})}
\nonumber \\
B=\frac{1}{2\pi_3}\sqrt{(\frac{J_1-J_2}{3}+\pi_3-\frac{\pi_2}{\sqrt{3}})
(\frac{J_1+2J_2}{3}-\pi_3+\frac{\pi_2}{\sqrt{3}})(\frac{2J_1+J_2}{3}+\pi_3-\frac{\pi_2}{\sqrt{3}})}
\label{eqb07}
\end{eqnarray}

$J_1$ and $J_2$ define the quadratic Casimir $q_2$ and cubic Casimir $q_3$.
Specifically,

\begin{eqnarray}
& & q_2=\sum_{\alpha}^{N_c^2-1}Q^{\alpha}Q^{\alpha}=
\frac{1}{3}(J_1^2+J_1J_2+J_2^2) \nonumber \\
& & q_3=\sum_{\alpha,\beta,\gamma}^{N_c^2-1}d_{\alpha\beta\gamma}
Q^{\alpha}Q^{\beta}Q^{\gamma}= \frac{1}{18}(J_1-J_2)
(J_1+2J_2)(2J_1+J_2) \label{eqb08}
\end{eqnarray}
with the help of the SU(2) symmetric tensor $d_{\alpha\beta\gamma}$.
we recall that
$d_{\alpha\beta\gamma}=d^{\alpha\beta\gamma}$ and
$d^{\alpha\beta\gamma}=d^{\beta\gamma\alpha}=d^{\gamma\alpha\beta}
=d^{\alpha\gamma\beta}=d^{\gamma\beta\alpha}=d^{\beta\alpha\gamma}$.

\noindent To carry some of the color integrations in the text, we
quote some useful identities involving $d$'s. For instance,

\begin{eqnarray}
& & \sum_{\alpha,\beta,\gamma}d^{\alpha\beta
\gamma}d^{\alpha\beta \gamma}=4q_3 \nonumber \\
& & \sum_{\alpha,\beta,\gamma}d^{\alpha\alpha\gamma}
d^{\beta\beta\gamma}=0 \label{eqb09}
\end{eqnarray}
and

\begin{eqnarray}
& & \sum_{\alpha,\beta,\gamma,\delta,\epsilon,
\zeta}d^{\alpha\gamma\epsilon}d^{\beta\gamma\epsilon} d^{\alpha
\delta\zeta}d^{\beta\delta\zeta}=2{q_3}^2  \nonumber \\
& & \sum_{\alpha,\beta,\gamma,\delta,\epsilon,
\zeta}d^{\alpha\beta\epsilon}d^{\gamma\delta\epsilon} d^{\alpha
\gamma\zeta}d^{\beta\delta\zeta}=-2q_3 \label{eqb10}
\end{eqnarray}

The SU(3) phase space measure is

\begin{eqnarray}
\lefteqn{dQ=c_R
d\phi_1d\phi_2d\phi_3d\pi_1d\pi_2d\pi_3dJ_1dJ_2\frac{\sqrt{3}}{48}J_1J_2(J_1+J_2)
} \nonumber \\
& \times \delta(\frac{1}{3}(J_1^2+J_1J_2+J_2^2)-q_2)
\delta(\frac{1}{18}(J_1-J_2) (J_1+2J_2)(2J_1+J_2)-q_3)
\label{eqb11}
\end{eqnarray}
with again $c_R$ set by the normalization of the color space volume to 1.

\begin{table}[!h]
\caption{Non-zero constants of $d^{\alpha\beta\gamma}$ for
SU$(3)$ } \label{tb1}
\begin{center}
\begin{tabular}{cccccccc}

$d^{118}$ & $d^{228}$ & $d^{338}$ & $d^{448}$ & $d^{558}$ &
$d^{668}$ & $d^{778}$ & $d^{888}$ \\ \hline $\frac{1}{\sqrt{3}}$ &
$\frac{1}{\sqrt{3}}$ & $\frac{1}{\sqrt{3}}$ &
$-\frac{1}{2\sqrt{3}}$ & $-\frac{1}{2\sqrt{3}}$ &
$-\frac{1}{2\sqrt{3}}$ & $-\frac{1}{2\sqrt{3}}$ &
$-\frac{1}{\sqrt{3}}$ \\ \\ $d^{146}$ & $d^{157}$ & $d^{247}$
& $d^{256}$ & $d^{344}$ & $d^{355}$ & $d^{366}$ & $d^{377}$ \\
\hline $\frac{1}{2}$ & $\frac{1}{2}$ & $-\frac{1}{2}$ &
$\frac{1}{2}$ & $\frac{1}{2}$ & $\frac{1}{2}$ & $-\frac{1}{2}$ &
$-\frac{1}{2}$

\end{tabular}
\end{center}
\end{table}

\section{Color integrations}
\renewcommand{\theequation}{B.\arabic{equation}}
\setcounter{equation}{0}

In this Appendix we explicit some of the color integrations
carried in the text. For SU(2) all color charge integrations can
be done analytically. The representation dependent $c_R$ is set to
give the normalization of the colored space volume to be 1. For
instance $c_R=1/2\pi \sqrt{q_2}$\cite{litim&manuel} by using $\int
dQ=1$. The SU(3) integrations cannot be done analytically given
the cubic nature of the constraint in the color measure. The
quadratic and cubic Casimirs are fixed by

\begin{eqnarray}
& & \int dQ q_2 = \int dQ \frac{1}{3}(J_1^2+J_1J_2+J_2^2) = q_2
\nonumber \\
& & \int dQ q_3 = \int dQ
\frac{1}{18}(J_1-J_2)(J_1+2J_2)(2J_1+J_2) = q_3 \label{eqc01}
\end{eqnarray}

\noindent{\bf One Color:}\\

\begin{equation}
\int dQ Q^{\alpha}=0 \label{eqc02}\,\,.
\end{equation}
\\
\\

\noindent{\bf Two Colors:}\\

\begin{equation}
\int dQ Q^{\alpha}Q^{\beta}=C_2\delta^{\alpha\beta}=
\frac{q_2}{N_c^2-1}\delta^{\alpha\beta} \label{eqc03}
\end{equation}

\noindent with $C_2=N_c$ for gluons and $C_2=1/2$ for quarks. For
SU$(3)$, the cubic Casimir $q_3=0$ for gluons and
$q_3=(N_c^2-1)(N_c^2-4)/4N_c$ for quarks. It is zero for SU(2).
\\
\\
\noindent{\bf Three Colors:}\\

\begin{eqnarray}
\sum_{\alpha,\beta,\gamma}d_{\alpha\beta\gamma}Q^{\alpha}Q^{\beta}
Q^{\gamma}=q_3=\frac{1}{4} \sum_{\alpha,\beta,\gamma}
d_{\alpha\beta\gamma}d^{\alpha\beta\gamma} = \int dQ q_3 =
\sum_{\alpha,\beta,\gamma} d_{\alpha\beta\gamma} \int dQ
Q^{\alpha}Q^{\beta}Q^{\gamma} \label{eqc05}
\end{eqnarray}
where we used (\ref{eqb08}) and (\ref{eqc01}) and the identity

\begin{equation}
\sum_{\alpha,\beta,\gamma}d^{\alpha\beta \gamma}d_{\alpha\beta
\gamma}=\frac{1}{N_c}(N_c^2-1)(N_c^2-4)=4q_3 \label{eqc04}
\end{equation}
Thus

\begin{equation}
\int dQ Q^{\alpha}Q^{\beta}Q^{\gamma}
=\frac{1}{4}d^{\alpha\beta\gamma} \label{eqc06}
\end{equation}
\\
\\

\noindent{\bf Four Colors:}\\

\begin{equation}
\int dQ Q^{\alpha}Q^{\beta}Q^{\gamma}Q^{\delta} = A ( \sum_{n}
d^{\alpha\beta n}d^{\gamma\delta n}+ \sum_{n}d^{\alpha\gamma
n}d^{\beta\delta n}+ \sum_{n}d^{\alpha\delta n}d^{\beta\gamma n}
)+B (\delta^{\alpha\beta}\delta^{\gamma\delta}+
\delta^{\alpha\gamma}\delta^{\beta\delta}+ \delta^{\alpha\delta}
\delta^{\beta\gamma}) \label{eqc07}
\end{equation}
For SU(2) $A=0$ and $B=q_2^2/15$. For SU(3) this integral is more
involved. However, for certain arrangements of charges, say
$Q^1,Q^2,Q^3,Q^8$ the integrations can be done and thus the
constants $A,B$ fixed. For that we need to undo integrals
involving $\pi's$ since $Q^8=\pi^2$.

The $\pi$ integrations are bounded~\cite{johnson,jeon&venugopalan}.
To carry them, it is best to change variables

\begin{equation}
x=\pi_3+\frac{\pi_2}{\sqrt{3}}, \quad
y=\pi_3-\frac{\pi_2}{\sqrt{3}} \label{eqc08}
\end{equation}
so that

\begin{equation}
d\pi_2d\pi_3=\frac{\sqrt{3}}{2}dxdy \label{eqc09}
\end{equation}
It is also useful to define

\begin{equation}
K_1=\frac{1}{3}(2J_1+J_2), \quad K_2=\frac{1}{3}(J_1+2J_2)
\label{eqc10}
\end{equation}
so that

\begin{equation}
K_2-K_1<x<K_1, \quad K_1-K_2<y<K_2 \label{eqc11}
\end{equation}
With these definitions, we can esily unwind some $\pi$
integrations. For instance

\begin{equation}
\int d\pi_1 d\pi_2 d\pi_3 = \frac{\sqrt{3}}{2}
\int_{K_2-K_1}^{K_1}dx\int_{K_1-K_2}^{K_2}dy
\int_{-\frac{1}{2}(x+y)}^{\frac{1}{2}(x+y)}d\pi_1=
\frac{\sqrt{3}}{4}J_1J_2(J_1+J_2) \label{eqc12}
\end{equation}
and

\begin{eqnarray}
\int d\pi_1 {\pi_2}^2d\pi_2 d\pi_3 & & = \frac{\sqrt{3}}{2}
\int_{K_2-K_1}^{K_1}dx\int_{K_1-K_2}^{K_2}dy \Big(
\frac{\sqrt{3}}{2}(x-y)\Big)^2
\int_{-\frac{1}{2}(x+y)}^{\frac{1}{2}(x+y)}d\pi_1 \nonumber \\
& & = \frac{1}{8}\frac{\sqrt{3}}{4}J_1J_2(J_1+J_2)
\frac{1}{3}(J_1^2+J_1J_2+J_2^2)\,\,. \label{eqc13}
\end{eqnarray}
This latter integral is directly related to

\begin{eqnarray}
\int dQ Q^8 Q^8 & & = c_R \int d\phi_1d\phi_2d\phi_3 \int
d\pi_1{\pi_2}^2d\pi_2d\pi_3 \int
dJ_1dJ_2\frac{\sqrt{3}}{48}J_1J_2(J_1+J_2) \nonumber \\
& & \times \delta(\frac{1}{3}(J_1^2+J_1J_2+J_2^2)-q_2)
\delta(\frac{1}{18}(J_1-J_2) (J_1+2J_2)(2J_1+J_2)-q_3) \nonumber \\
& & = \int dQ \frac{1}{8}\frac{1}{3}(J_1^2+J_1J_2+J_2^2) = \int dQ
\frac{1}{8}q_2 = \frac{1}{N_c^2-1}q_2 \label{eqc14}
\end{eqnarray}
after using (\ref{eqb11}) and (\ref{eqc01}). The
result agrees with (\ref{eqc03}). Similarly,

\begin{eqnarray}
\int d\pi_1 {\pi_2}^3d\pi_2 d\pi_3 & & = \frac{\sqrt{3}}{2}
\int_{K_2-K_1}^{K_1}dx\int_{K_1-K_2}^{K_2}dy \Big(
\frac{\sqrt{3}}{2}(x-y)\Big)^3
\int_{-\frac{1}{2}(x+y)}^{\frac{1}{2}(x+y)}d\pi_1 \nonumber \\
& & = -\frac{\sqrt{3}}{40}\frac{\sqrt{3}}{4} J_1J_2 (J_1+J_2)
\frac{1}{18}(J_1-J_2) (J_1+2J_2)(2J_1+J_2) \label{eqc15}
\end{eqnarray}
is related to

\begin{eqnarray}
\int dQ Q^8 Q^8 Q^8 & & = c_R \int d\phi_1d\phi_2d\phi_3 \int
d\pi_1{\pi_2}^3d\pi_2d\pi_3 \int
dJ_1dJ_2\frac{\sqrt{3}}{48}J_1J_2(J_1+J_2)  \nonumber \\
& & \times \delta(\frac{1}{3}(J_1^2+J_1J_2+J_2^2)-q_2)
\delta(\frac{1}{18}(J_1-J_2) (J_1+2J_2)(2J_1+J_2)-q_3) \nonumber \\
& & = -\frac{\sqrt{3}}{40} \int dQ \frac{1}{18}(J_1-J_2)
(J_1+2J_2)(2J_1+J_2) \nonumber \\
& & = -\frac{\sqrt{3}}{40} \int dQ q_3 =
-\frac{1}{4}\frac{1}{\sqrt{3}}\frac{3}{10} q_3 =
\frac{1}{4}d^{888} \label{eqc16}
\end{eqnarray}
after using $q_3=(N_c^2-1)(N_c^2-4)/4N_c=10/3$ for
SU$(3)$. This result is in agreement with (\ref{eqc06}).
A similirar reasoning, yields

\begin{equation}
A+3B=\frac{3}{80}q_2^2 \label{eqc17}
\end{equation}
for the $A,B$ constants in (\ref{eqc07}).
\\
\\

\noindent{\bf Five Colors:}\\

\begin{eqnarray}
\lefteqn{\int dQ
Q^{\alpha}Q^{\beta}Q^{\gamma}Q^{\delta}Q^{\epsilon} =
\frac{3}{560}q_2 q_3 \Big(
\delta^{\alpha\beta}d^{\gamma\delta\epsilon}
+\delta^{\alpha\gamma}d^{\beta\delta\epsilon}
+\delta^{\alpha\delta}d^{\beta\gamma\epsilon}
+\delta^{\alpha\epsilon}d^{\beta\gamma\delta} } \nonumber \\
& +\delta^{\beta\gamma}d^{\alpha\delta\epsilon}
+\delta^{\beta\delta}d^{\alpha\gamma\epsilon}
+\delta^{\beta\epsilon}d^{\alpha\gamma\delta}
+\delta^{\gamma\delta}d^{\alpha\beta\epsilon}
+\delta^{\gamma\epsilon}d^{\alpha\beta\delta}
+\delta^{\delta\epsilon}d^{\alpha\beta\gamma} \Big) \label{eqc18}
\end{eqnarray}
with

\begin{equation}
\frac{3}{560} q_2 q_3 = \frac{1}{7}\frac{1}{8}q_2\frac{3}{10} q_3
= \frac{1}{7}C_2 \label{eqc19}
\end{equation}
\\
\\

\noindent{\bf Six Colors:}\\

\begin{eqnarray}
\int dQ
Q^{\alpha}Q^{\beta}Q^{\gamma}Q^{\delta}Q^{\epsilon}Q^{\zeta} & & =
A \Big( d^{\alpha\beta\gamma}d^{\delta\epsilon\zeta}
+d^{\alpha\beta\delta}d^{\gamma\epsilon\zeta}
+d^{\alpha\beta\epsilon}d^{\gamma\delta\zeta}
+d^{\alpha\beta\zeta}d^{\gamma\delta\epsilon}
+d^{\alpha\gamma\delta}d^{\beta\epsilon\zeta} \nonumber \\
& &  +d^{\alpha\gamma\epsilon}d^{\beta\delta\zeta}
+d^{\alpha\gamma\zeta}d^{\beta\delta\epsilon}
+d^{\alpha\delta\epsilon}d^{\beta\gamma\zeta}
+d^{\alpha\delta\zeta}d^{\beta\gamma\epsilon}
+d^{\alpha\epsilon\zeta}d^{\beta\gamma\delta} \Big) \nonumber \\
& & + B \Big(
\delta^{\alpha\beta}\delta^{\gamma\delta}\delta^{\epsilon\zeta} +
\delta^{\alpha\beta}\delta^{\gamma\epsilon}\delta^{\delta\zeta} +
\delta^{\alpha\beta}\delta^{\gamma\zeta}\delta^{\delta\epsilon} +
\delta^{\alpha\gamma}\delta^{\beta\delta}\delta^{\epsilon\zeta} +
\delta^{\alpha\gamma}\delta^{\beta\epsilon}\delta^{\delta\zeta}
\nonumber \\
& & +
\delta^{\alpha\gamma}\delta^{\beta\zeta}\delta^{\delta\epsilon} +
\delta^{\alpha\delta}\delta^{\beta\gamma}\delta^{\epsilon\zeta} +
\delta^{\alpha\delta}\delta^{\beta\epsilon}\delta^{\gamma\zeta} +
\delta^{\alpha\delta}\delta^{\beta\zeta}\delta^{\gamma\epsilon} +
\delta^{\alpha\epsilon}\delta^{\beta\gamma}\delta^{\delta\zeta}
\nonumber \\
& & +
\delta^{\alpha\epsilon}\delta^{\beta\delta}\delta^{\gamma\zeta} +
\delta^{\alpha\epsilon}\delta^{\beta\zeta}\delta^{\gamma\delta} +
\delta^{\alpha\zeta}\delta^{\beta\gamma}\delta^{\delta\epsilon} +
\delta^{\alpha\zeta}\delta^{\beta\delta}\delta^{\gamma\epsilon} +
\delta^{\alpha\zeta}\delta^{\beta\epsilon}\delta^{\gamma\delta}
\Big) \nonumber \\ \label{eqc20}
\end{eqnarray}
with the $A,B$ constants tabulated below.

\begin{table}[!h]
\caption{Constants A and B in the integration of six color charges
} \label{tb2}
\begin{center}
\begin{tabular}{ccc}
\hline

$  $ & $ A $ & $ B $ \\
\hline
$SU(2)$ & $   $ & $ \frac{1}{105}q_2^3 $ \\
$SU(3)$ & $ -\frac{9}{8!}q_2^3+\frac{27}{2}\frac{1}{7!}q_3^2$ & $
\frac{85}{2}\frac{1}{8!}q_2^3-\frac{6}{8!}q_3^2$ \\
\hline

\end{tabular}
\end{center}
\end{table}

\end{document}